\documentclass[authoryear]{elsarticle}
\usepackage{graphicx}
\usepackage{natbib,rotfloat,count1to,relsize,amsmath}
\usepackage{txfonts}
\newcommand\sun{\odot}%
\def\apj{ApJ }%
\def\apjl{ApJ }%
\def\ao{Appl.~Opt. }%
\def\apss{Ap\&SS }%
\def\aap{A\&A }%
\begin{document}

\begin{frontmatter}

\title{The thermal structure and the location of the snow line in the protosolar nebula: axisymmetric models with full 3-D radiative transfer}

\author[uu]{M.~Min}
\ead{M.Min@uu.nl}
\author[mpia]{C.P.~Dullemond}
\author[uva]{M.~Kama}
\author[uva,run]{C.~Dominik}

\address[uu]{Astronomical institute Utrecht, Utrecht University, P.O. Box 80000, NL-3508 TA Utrecht, The Netherlands}
\address[mpia]{Max Planck Institut f\"ur Astronomie, K\"onigstuhl 17, 69117 Heidelberg, Germany}
\address[uva]{Astronomical Institute `Anton Pannekoek', Science Park 904, NL-1098 XH Amsterdam, The Netherlands}
\address[run]{Department of Astrophysics/IMAPP, Radboud University Nijmegen, P.O. Box 9010, 6500 GL Nijmegen, The Netherlands}

\begin{abstract}
The precise location of the water ice condensation front (Ôsnow lineÕ) in 
the protosolar nebula has been a debate for a long time. Its importance 
stems from the expected substantial jump in the abundance of solids beyond 
the snow line, which is conducive to planet formation, and from the higher 
ÔstickinessÕ in collisions of ice-coated dust grains, which may help the process 
of coagulation of dust and the formation of planetesimals. In an optically thin 
nebula, the location of the snow line is easily calculated to be around 3 AU,
subject to brightness variations of the young Sun. However, in its first 5 to 10
million years, the solar nebula was optically thick, implying a smaller snowline 
radius due to shielding from direct sunlight, but also a larger radius because of
viscous heating. Several models have attempted to treat these opposing 
effects. However, until recently treatments beyond an approximate 1+1D
radiative transfer were unfeasible. We revisit the problem with a fully
self-consistent 3D treatment in an axisymmetric disk model, including a 
density-dependent treatment of the dust and ice sublimation. We find that the 
location of the snow line is very sensitive to the opacities of the dust grains and 
the mass accretion rate of the disk. We show that previous approximate 
treatments are quite efficient at determining the location of the snow line if the 
energy budget is locally dominated by viscous accretion. Using this result we 
derive an analytic estimate of the location of the snow line that compares very well 
with results from this and previous studies. Using solar abundances of the 
elements we compute the abundance of dust and ice and find that the expected jump 
in solid surface density at the snow line is smaller than previously assumed. 
We further show that in the inner few AU the refractory species are also partly 
evaporated, leading to a significantly smaller solid state surface density in the 
regions where the rocky planets were formed.
\end{abstract}


\end{frontmatter}

\section{Introduction}

The efficiency of the formation of rocky planets and the cores of gas giant
planets depends sensitively on the amount of solids present in each location
of the solar nebula. In most models of planet formation the solar nebula has
been assumed to resemble the structures proposed by \citet{1977Ap&SS..51..153W} 
and \citet{1981IAUS...93..113H}. In this model, and subsequent incarnations
of it, the surface density of the gas is approximately given by
\begin{equation}
\label{eq:gas mmsn}
\Sigma^{\mathrm{MMSN}}_{\mathrm{gas}}(r) = 1700 \left(\frac{R}{1\;\mathrm{AU}}\right)^{-3/2}\quad \mathrm{gram/cm}^2
\end{equation}
\citep[see][]{2006plfo.book..129T}
where the acronym MMSN stands for ``minimum mass solar nebula'', a name by
which this model is often identified. In the inner regions (small values of $R$) of
the nebula the dust grains are refractory in nature, while beyond some
radius $R_{\mathrm{ice}}$ water ice is present, most likely in the form of
ice mantels surrounding the refractory grains. In the Hayashi MMSN model
this location is assumed to be $R_{\mathrm{ice}}=2.7$ AU, which is
consistent for temperatures expected in an optically thin nebula. Often this
water ice sublimation boundary radius is called the `snow line'. Since ices may
contain a considerable mass compared to the mass in refractory grains, the
total surface density was assumed to jump at $R=R_{\mathrm{ice}}$:
\begin{equation}
\label{eq:mmsn}
\Sigma^{\mathrm{MMSN}}_{\mathrm{solid}}(r) = 7.1 F_{\mathrm{ice}} \left(\frac{R}{1\;\mathrm{AU}}\right)^{-3/2}\quad \mathrm{gram/cm}^2
\end{equation}
with 
\begin{equation}
F_{\mathrm{ice}} =\left\{ \begin{matrix}
1, & \hbox{for} & R< R_{\mathrm{ice}} \\
4.2, & \hbox{for} & R> R_{\mathrm{ice}} 
\end{matrix}\right.
\end{equation}
\citep{2006plfo.book..129T}

This model serves in many planet formation models to compute the midplane
gas and solid density and the isolation mass of planetary embryos. However,
it has been shown that the outcomes of planet formation models depends
critically on the value of $R_{\mathrm{ice}}$ and the strength of the ice
jump $F_{\mathrm{ice}}$ \citep[see e.g.][]{2009A&A...501.1139M}. 
Also dust coagulation and planet formation and migration models are very sensitive to
these parameters \citep[e.g.][]{2008A&A...480..859B, 2008ApJ...685..584I}. 
Simply taking the values to be $R_{\mathrm{ice}}=2.7$\,AU and
$F_{\mathrm{ice}}=4.2$ (for $R>R_{\mathrm{ice}}$) is probably too simplistic
and may lead to major errors in predictions of the progress of the planet
formation process.

Besides the amount of solid mass available also the strength of the bonds between dust grains is of importance to their aggregation behavior. The presence of ice coatings on grains can significantly increase their 'stickiness', allowing grains to coagulate more easily. Perhaps even more importantly the bonds formed are stronger preventing aggregates that formed to be destroyed again by collisions.

Over the last decade or so, there has been enormous progress in astronomy in
the understanding of the structure and evolution of `protoplanetary disks',
i.e.~the dust+gas disks surrounding many very young stars. These disks very
likely give a good impression of what our own solar nebula looked like 4.567
billion years ago, and the knowledge gained in that field can help getting a
better handle on the structure of the solar nebula and the location of the
ice condensation front. Indeed, modeling tools that were originally meant
for modeling protoplanetary disks and comparing model predictions to
observations have been used to make models of the solar nebula and the snow
line. For instance, the 1+1D disk structure model of \citet{2000A&A...358..378H}, 
in which the equations are first solved vertically (1D)
and then connected radially (1+1D), has been used by \citet{2001ApJ...554..391H} 
to study water in the solar nebula, in
particular concerning the issue of the D/H ratio in meteorites. \citet{2005ApJ...627L.153D} 
made a model of the surface density of the solar
nebula disk as a function of radial coordinate and accretion rate. He used
this model to study the ice condensation front \citep{2005ApJ...620..994D}. 
This model includes both the effect of irradiation of the disk by
the young sun and the heating of the disk near the midplane through viscous
dissipation of potential energy (i.e.\ accretional heating). Like in the
Hersant study, these models are 1+1D models, i.e.\ they are 1-D vertical
models of density and temperature as a function of $z$. The Davis models
include radiative transfer using a variable eddington factor
method. \citet{2006ApJ...640.1115L} also modeled this, using an updated 
version of the \citet{1997ApJ...490..368C} two-layer disk model, i.e.\ slightly simpler than the 1+1D
modeling of Davis. Recently, \citet{2009Icar..200..672D} combined complex models of ice formation and viscous evolution of the disk with a simple radiative transfer approximation to compute the evolution of the solid surface density.

While these modeling efforts have improved the understanding of the
distribution of ices in the solar nebula substantially, they still rely on
rather dramatic simplifications of the treatment of radiative transfer.  The
reason for this is very understandable: it is quite a numerical challenge to
model multi-dimensional radiative transfer in protoplanetary disks that are
very optically thick and actively accreting (and hence producing heat near
the midplane). However, in recent years multi-dimensional radiative transfer
tools, in particular those based on the Monte Carlo method, have improved
dramatically in speed. Extreme optical depths are now not necessarily a
problem anymore. We have developed a code, MCMax, that can now handle
extreme optical depths efficiently, even if many of the
photon packages originate from the optically thickest regions of the
disk \citep{2009A&A...497..155M}. In addition, we have adjusted the code to fully take into account the density dependend sublimation state of both the refractory as well as the icy material \citep{2009A&A...506.1199K}.
It is therefore a right time to revisit the 
problem of the snow line with the full force of multi-dimensional radiative transfer and investigate
(a) where is the ice line as a function of time and (b) what is the gas and
solid surface density distribution in the disk as a function of time. This
is the goal of this paper.

\section{Model}

Our model combines the radiative transfer equipment and basic disk model
setup described in \citet{2009A&A...497..155M}. We add here: a self-consistent
and automatic computation of the radial structure, viscous heating of the disk by accretion
and a self-consistent treatment of water ice and dust evaporation.

Right at this point we already note that we will replace `time' $t$ with
`accretion rate' $\dot M$ and study only the inner 20\,AU of the disk. The
reason is that the long term evolution of protoplanetary disks depends very
strongly on processes such as photoevaporation of the {\em outer} disk
regions \citep[see e.g.][]{2009ApJ...690.1539G}. The inner disk regions, say inward of 20 AU, are regulated
entirely by the `feeding' of matter by the outer disk. How this feeding
depends on time requires a detailed study of the complex processes happening
at large radii. But given a recipe of feeding, i.e.\ an accretion rate as a
function of time $\dot M(t)$, the inner disk can be assumed to be entirely
determined. In particular if we avoid issues such as ionization degree and
possible instabilities in the disk \citep[see e.g.][]{2006ApJ...648..484H} 
and assume that no mass pile-up can
happen anywhere in the inner disk (this assumption must be relaxed in future
work), then the stationary disk structure inward of 20 AU is entirely
determined by the stellar parameters and the accretion rate $\dot M$. For
any given star our model thus becomes a {\em 1-parameter model}. So instead
of having $t$ as our parameter, we will have $\dot M$ as our parameter.
Generally, a high mass accretion rate corresponds to early times in the evolution
of the solar nebula, while low mass accretion rates correspond to later times.
Alternatively, in the episodic accretion scenario, low mass accretion
rates represent the typical state, which is punctuated by relatively short
episodes with high mass accretion rates.

The main improvement of our study over earlier work is that we treat the
flow of radiation through the disk in a fully self-consistent manner,
assuming that dust opacities are everywhere dominant over the gas opacities.
Our model is axially symmetric, i.e.\ all variables depend on radial
coordinate $r$ and vertical coordinate $z$, but do {\em not} depend on
longitudinal coordinate $\phi$. The movement of photons is, however, fully
3-D, i.e.\ a photon can move also outside of the $r,z$-plane and has all the
3-D freedom of motion. In our Monte Carlo scheme \citep[based on][]{2001ApJ...554..615B} we
split the total input luminosity $L$ into $N$ `photon packages', each
carrying a luminosity $L_{\mathrm{package}}=L/N$. The total luminosity
includes the stellar luminosity $L_{*}$ and the accretional heating
luminosity $L_{\mathrm{visc}}$, i.e.~$L=L_{*}+L_{\mathrm{visc}}$. As these
photon packages travel through the disk they update the local temperature of
the dust according to the scheme by \citet{2001ApJ...554..615B}. Since our
model is axisymmetric, these updates have to only be done in 2-D, i.e.\ the
temperature only depends on (and will only be updated in) $r,z$, no matter
what is the longitudinal location $\phi$ of the photon package. 
So while the radiative transfer is 3-D, the resulting structure is 2-D.

The main difference between models doing radiative transfer in 1+1D and our model doing full 3D is in the radial energy diffusion. In the 1+1D model the energy from the star is intercepted by the disk and further diffusion is assumed to be vertical. This approximation is expected to be accurate for the regions of the disk close to the midplane where the very high densities do not allow much radial diffusion. However, radial energy diffusion is important in the surface layers and, more importantly, in the innermost regions of the disk, directly illuminated by the central star. Our model naturally takes this into account and allows us to see in which parts of the disk the vertical diffusion approximation is valid. Also, 3D radiative transfer is the only way to properly treat the complex shape of the dust condensation front close to the star \citep[see also][]{2009A&A...506.1199K}.

For the central star in our model we take the Sun, i.e. the mass $M_\star=M_{\sun}$, the luminosity $L_\star=L_{\sun}$, and the stellar radius $R_\star=R_{\sun}$. The spectrum of the Sun is approximated by a blackbody of 5777\,K. Although the parameters of the early Sun were likely different from this, we do not expect much difference on the results in this paper.

\subsection{Treatment of viscous accretion}

We assume that viscous heating is described by the so-called $\alpha$-disk model \citep{1973A&A....24..337S}. In this model it is assumed that the energy produced locally by viscous heating is proportional to the local gas pressure through the proportionality constant $\alpha$. The total energy locally created per unit volume is then given by
\begin{equation}
\label{eq:gamma}
\Gamma_\mathrm{viscous}=\frac{9}{4}\alpha P(z)\Omega(R),
\end{equation}
where $P(z)$ is the gas pressure, and $\Omega(R)$ is the Keplerian angular velocity at radius $R$ from the central star.
If we define a uniform accretion rate $\dot{M}$ throughout the disk we have for the total energy released per unit surface area of the disk per unit time at a radius $R$ from the Sun is
\begin{equation}
\label{eq:tot energy}
F_\mathrm{viscous}=\frac{3GM_\star\dot{M}}{4\pi}R^{-3}\left[1-\left(\frac{R_\star}{R}\right)^{1/2}\right],
\end{equation}
where $G$ is the gravitational contstant, and $M_\star$ and $R_\star$ are the mass and radius of the central star.
Combining Eqs.~\ref{eq:gamma} and \ref{eq:tot energy} we can solve for the surface density of the disk for any given value of $\alpha$. In this paper we will consider two values of $\alpha$, $0.1$ and $0.01$. From Eq.~(\ref{eq:tot energy}) it is clear that the total amount of energy released is independent of the structure of the disk. This speeds up convergence since there is no feedback from the disk structure on the total energy that is released.

We basically treat the energy released by viscous stress in the same way as the energy released by the central star. Thus a fraction of the photon packages released in the Monte Carlo radiative transfer process is now released from the inner regions of the disk. Where previous studies that take into account radiation from viscous stress rely on assumptions on the radiative transfer in the optically thick inner regions, we treat the radiative transfer fully consistent in the framework of Monte Carlo radiative transfer as outlined by \citet{2009A&A...497..155M}. The total accretional heating luminosity of the disk is given by the integral of Eq.~(\ref{eq:tot energy}) from the stellar surface, $R_\star$, to the outer radius, $R_\mathrm{out}$. Assuming $R_\mathrm{out}>>R_\star$ this leads to
\begin{equation}
L_\mathrm{visc}=\frac{1}{2}\,\frac{GM_\star\dot{M}}{R_\star}.
\end{equation}
Note that a large fraction of this energy is most likely released inside the dust sublimation radius and thus can easily escape the disk. This energy does not influence the dust temperature and is thus not taken into account in our computation. In our method the energy produced through viscous heating is directly deposited from the gas onto the dust grains. This assumes a strong coupling between gas and dust, which is an appropriate assumption in the regions where the densities of both gas and dust are high. In the regions where a large fraction of the dust is evaporated due to high temperatures, we limit the amount of energy deposited on the grains by viscous heating by taking into account the number of collisions between the gas and the dust and assuming a gas molecule can at maximum transfer its total kinetic energy onto the grain in such a collision. This limits the total energy that can be transferred to the dust per unit time. We find that indeed in the optically thick regions this limit exceeds the available energy by many orders of magnitude. However, in the inner regions, this limit prevents the small amount of dust grains available to be overloaded with accretional energy from the dense gas surrounding it. The dust grains are assumed to radiate the energy they absorbed with a wavelength distribution according to the local dust temperature after which the regular procedure of Monte Carlo radiative transfer is proceeded.

In the dusty regions of the disk the optical depth from the midplane, where most energy is released, to the outside is extremely large. Therefore, a photon package undergoes a large number of interaction steps before it leaves the system (on the order of several million interactions). If all these interaction steps would be computed separately, the computation time required would be on the order of a minute per photon package, putting strong constraints on the number of photon packages that can be used. 
\citet{2009A&A...497..155M} introduced an analytical solution for the random walk regime of the radiative transfer that can be applied in the optically thick regions and reduces computation times by orders of magnitude.  This algorithm therefore allows the use of a large number of photon packages and thus to obtain proper statistics everywhere.

\begin{table*}[!tbp]
\begin{center}
\linespread{1.3}%
\selectfont
\begin{tabular}{l|cccc|c|c|c}
				& Silicates 	& FeS		& C-dust		& Water ice 	& Gas to			& $F_\mathrm{ice}$	& $\kappa_R$\\
				&			&			&			&			& solid ratio		&				& [cm$^2$/g]\\
\hline
only CO		& 47\%		& 14\%		& -			& 39\%		& 97					& 1.63 			& 572\\
\small{($w=0$)}		&			&			&			&			&					&				&\\
only C-dust	& 24\%		& 7\%		& 20\%		& 49\%		& 49					& 1.97 			& 544\\
\small{($w=1$)}		&			&			&			&			&					&				&\\
mixture		& 32\%		& 10\%		& 13\%		& 45\%		& 66					& 1.84 			& 508\\
\small{($w=0.5$)}		&			&			&			&			&					&				&\\
\end{tabular}
\linespread{1}%
\selectfont
\end{center}
\caption{Parameters of the three different dust models considered. The composition is given as mass fractions. The gas to solid ratio is the ratio between the mass in gas and the total available mass to form solids. $F_\mathrm{ice}$ is the fractional increase in the mass available to form solids when water ice can condense. The value of the Rosseland mean opacity, $\kappa_R$, is taken at $160\,$K is computed considering all dust species including water ice.}
\label{tab:abundances}
\end{table*}

\subsection{The dust and ice composition and opacities}

In order to estimate the abundances of the various dust components we use here the Solar abundances of the elements as determined by \citet{1998SSRv...85..161G}. The dust mixture is then constructed as follows:
\begin{itemize}
\item Since iron sulfide is the dominant form of (solid) sulfur in meteorites in the solar system it is first assumed that all the available sulfur is in the form of FeS. This takes approximately half the available number of Fe-atoms.
\item Then we assume all Mg and Si go into silicates together with the remaining Fe-atoms. This creates silicates with an average composition of roughly MgFe$_{0.5}$SiO$_{3.5}$. We assume this to be in the form of amorphous silicates. This takes approximately 15\% of the available O-atoms.
\item Next we put a fraction $w$ of the available C-atoms in the form of carbonaceous dust grains. The remaining C-atoms are assumed to form CO. This takes away up to $\sim 50$\% of the available number of O-atoms depending on the value of $w$.
\item The remaining O-atoms are assumed to form water ice everywhere in the disk where the temperature allows it.
\end{itemize}
From the above procedure we also can compute the gas to dust ratio. 
We find that the resulting gas-to-dust ration can deviate up to a factor of 2 from the canonical value 100.
The question remains what the most realistic value of the carbon abundance in dust, $w$, is. We take as a standard the value $w=0.5$, based on in-situ measurements of the dust in comet Halley \citep{1987A&A...187..859G}. However, the carbon fraction in dust could have been higher in the early Solar nebula since the interstellar solid carbon abundance is much higher. Also, in the inner regions, at temperatures
above $\sim1000$\,K, carbon grains will be destroyed by combustion so lower values of $w$ are to be expected. The dust/ice/gas mixture resulting from this procedure for different choices of $w$ are summarized in Table~\ref{tab:abundances}.

In order to compute the opacities of the dust species as a function of wavelength we need to know the refractive indices of the grains. For the silicates we use the measurements by \citet{1995A&A...300..503D} and \citet{1996A&A...311..291H}, for the iron sulfide we take the laboratory data from \citet{1994ApJ...423L..71B}, for the carboneceous material we take the measurements from \cite{1993A&A...279..577P}, and for the water ice we take the data from \citet{1984ApOpt..23.1206W}. In order to convert the refractive index data to opacities we use the method by \citet{2005A&A...432..909M} to simulate irregularly shaped particles. The particle sizes are assumed to follow the so-called MRN distribution \citep{1977ApJ...217..425M}. We thus assume that no significant grain growth has occurred with respect to the grains in the interstellar medium. We will come back to this assumption in the discussion of the results.

\subsection{Treatment of evaporation and condensation}

The temperature at which a species evaporates depends on the vapor pressure of the material. We treat the evaporation and condensation of the dust species following \citet{2009A&A...506.1199K}. In order to converge the structure of the inner regions \citet{2009A&A...506.1199K} had to put constraints on the vertical and radial gradients of the fraction of the material in the gas phase. This is to avoid instabilities caused by the sudden jump in opacity at the inner edge of the disk. Since the ice is located in the inner regions of the disk, the ill conditioned convergence is no issue here. Therefore, for the water ice we do not put any constraints on the gradients of the condensed fraction. It turns out that convergence is reached relatively fast and no instabilities like those reported by \citet{2009A&A...506.1199K} are found for the ice sublimation zone. Although the density structure does not change significantly anymore after $\sim30$ iterations, we ran all models up to 100 iterations to make sure also the slowly changing structures are converged.

To simplify the evaporation of the refractory dust species we take the evaporation parameters for amorphous olivine for the entire dust mixture of silicates, iron sulfide and carboneceous material. For the iron sulfide it might be argued that this is not a correct assumption since the evaporation temperature of FeS is around 680\,K. However, we assume that above this temperature the iron condenses as metallic iron, contributing similarly to the opacity of the grains. The evaporation characteristics of both ice and olivine are based on sublimation data from \citet{1994ApJ...421..615P}.

\subsection{Convection}
\label{sec:convection}

In the implementation of radiative transfer and the computation of heat balance in the disk we employ, the effect of convection is not taken into account. However, we will show that convection is an important way of transporting energy generated in the midplane of the disk by accretion to the surface. Therefore, we have implemented a rough treatment of convection in the radiative transfer code to test the effects on our results. In the computations below, convection is not taken into account unless stated otherwise.

\begin{figure}[!tb]
\centerline{\resizebox{0.9\hsize}{!}{\includegraphics{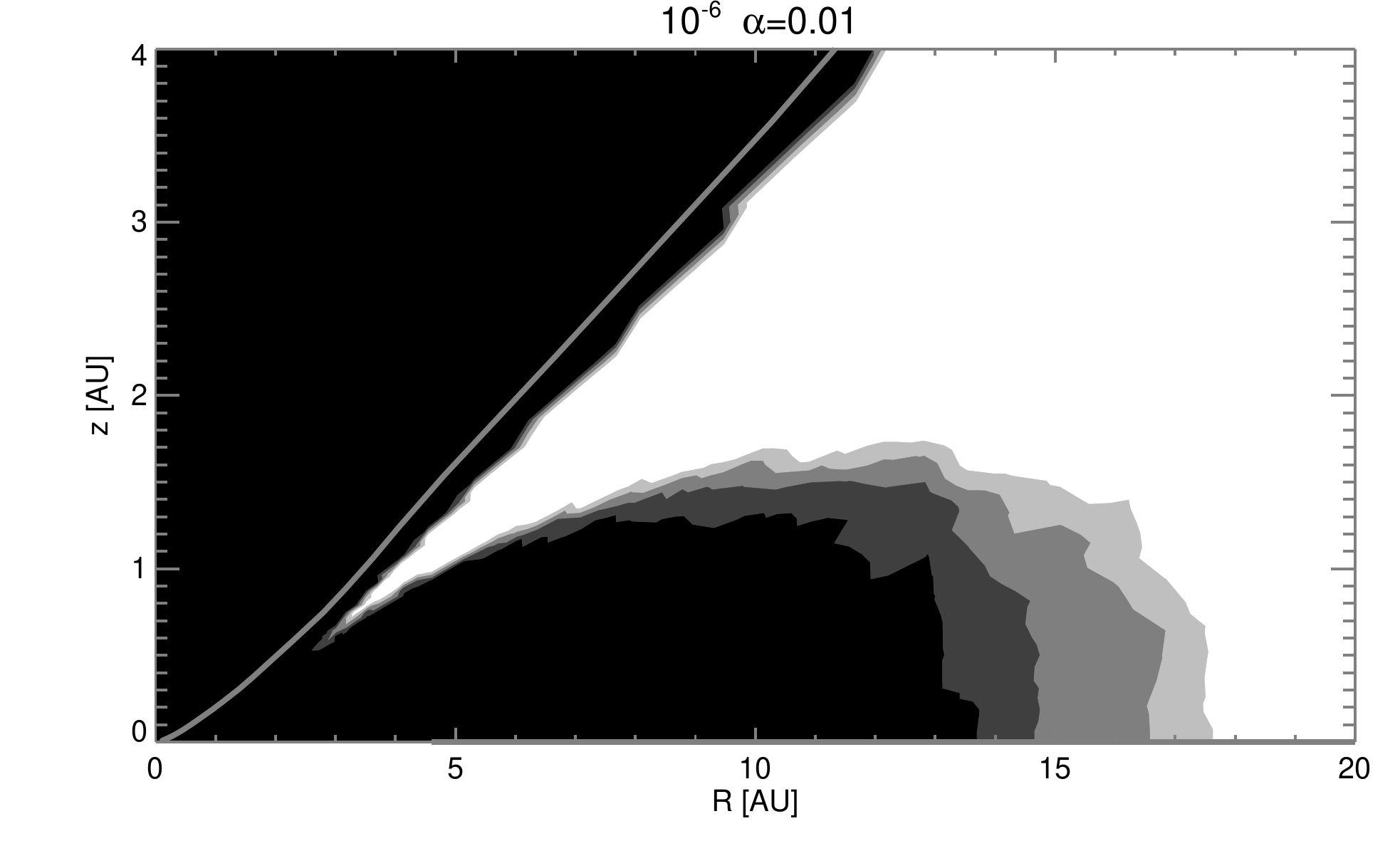}\includegraphics{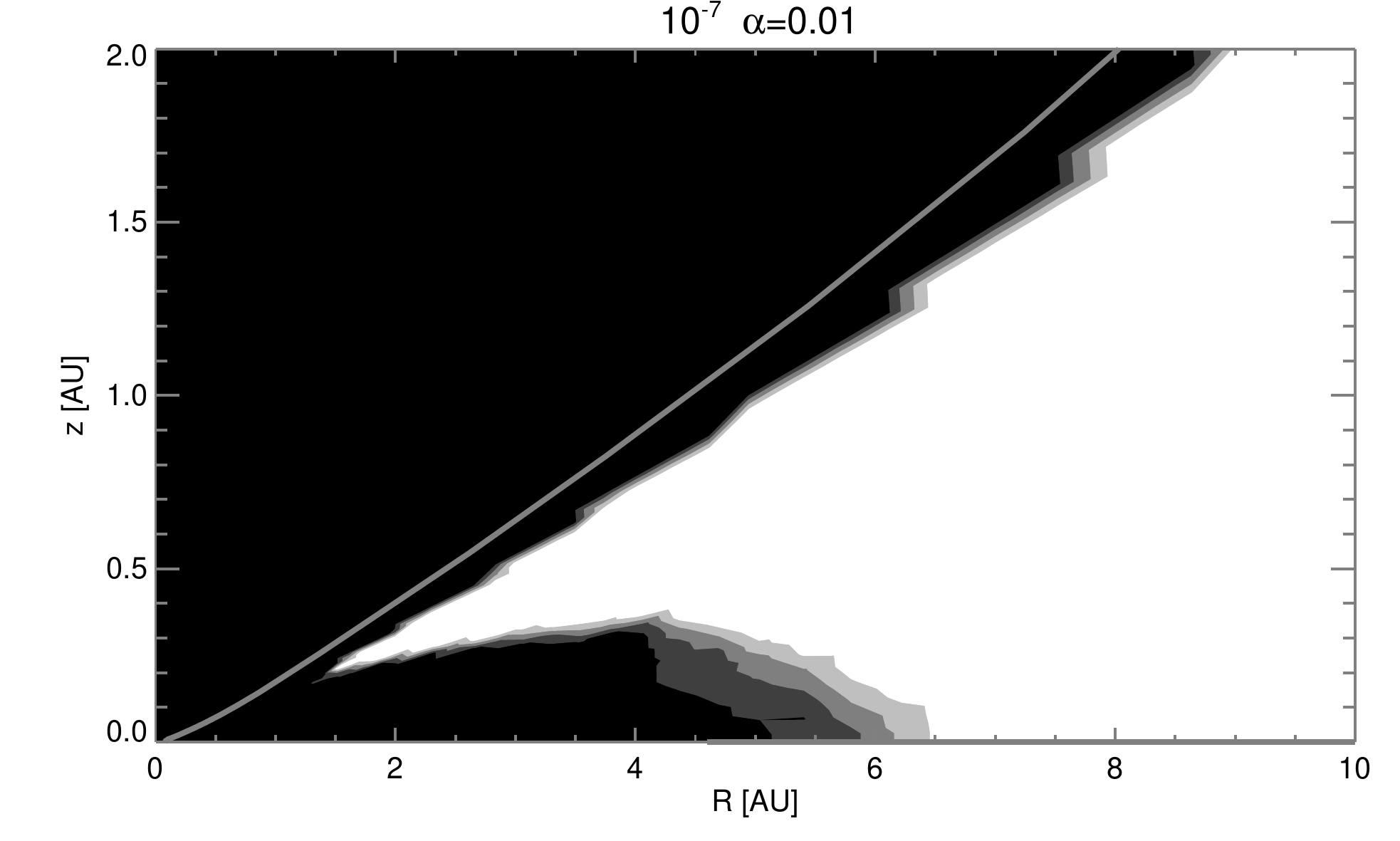}}}
\centerline{\resizebox{0.9\hsize}{!}{\includegraphics{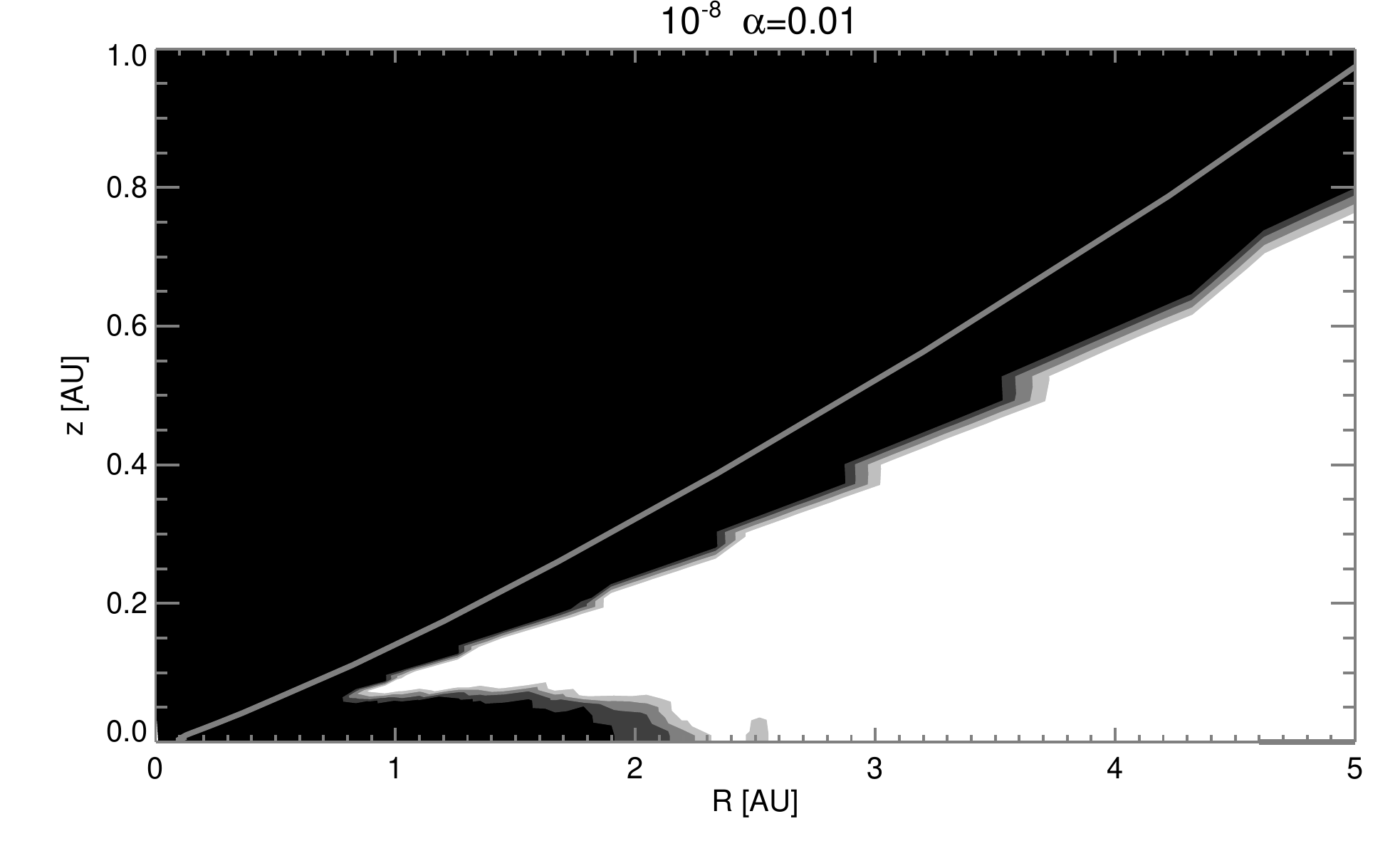}\includegraphics{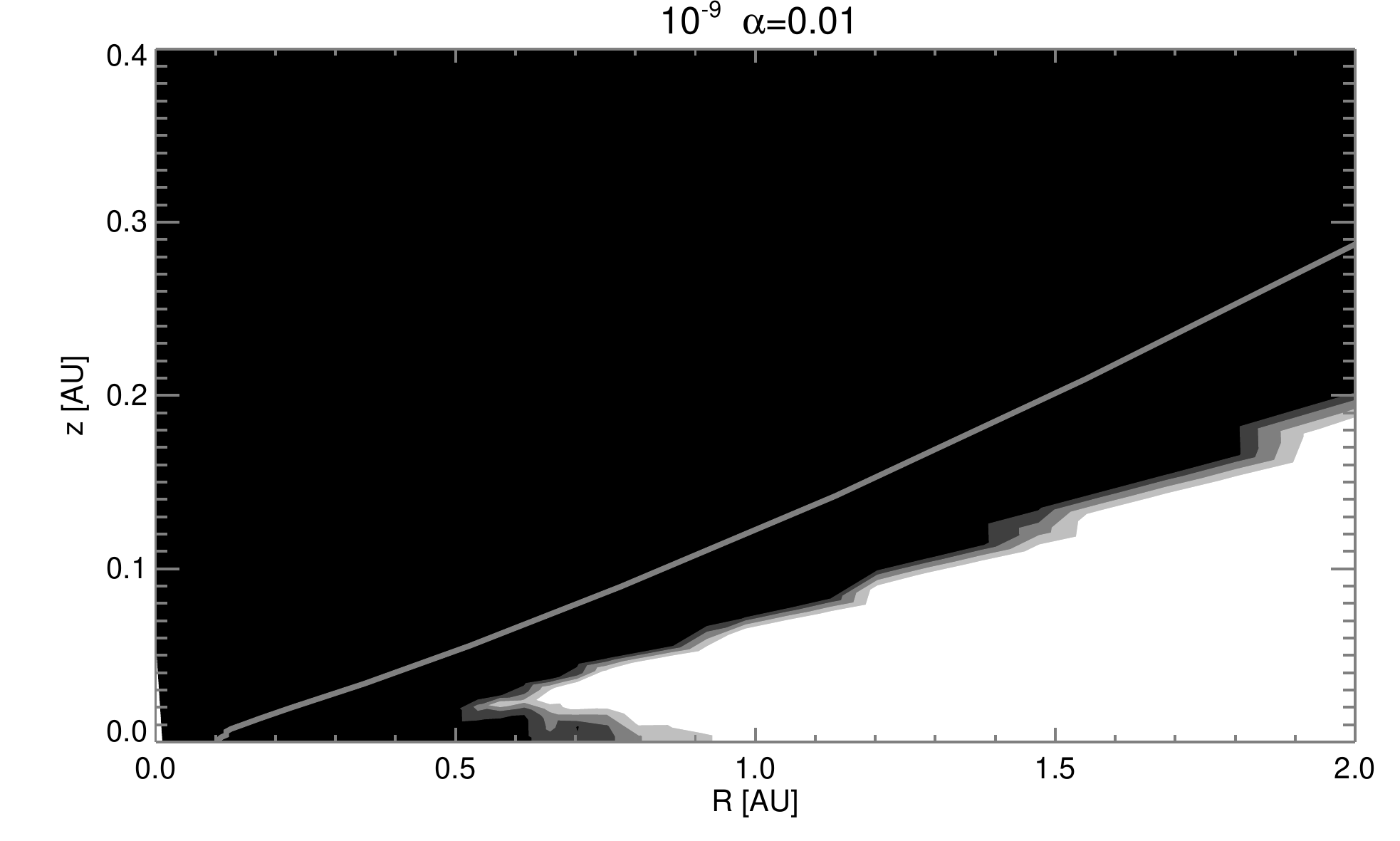}}}
\caption{The areas in the disk where water ice is present for various mass accretion rates. The case in which half of the available C atoms are in the form of solid carbon is presented for a value of the viscosity parameter $\alpha=0.01$. Increasing intensity corresponds to the region where up to 25\%, 50\%, 75\% and 100\% of the available ice is condensed. The white line corresponds to the location where the radiation from the star hits an optical depth of one.} 
\label{fig:snow lineCarbon0.5a0.01}
\end{figure}

Convection becomes an important way of transporting heat when the radiative temperature gradient becomes too large. According to the Schwarzschild criterion for convection this happens when
\begin{equation}
\label{eq:convection}
\frac{d\ln T}{d\ln P}>\nabla_\mathrm{ad},
\end{equation}
where $T$ and $P$ are the gas density and pressure, and $\nabla_\mathrm{ad}$ is the adiabatic limit which we take to be that of a diatomic gas: $\nabla_\mathrm{ad}=2/7$. Our rough implementation to correct the obtained temperature structure for convection is simply as follows. We start at the surface of the disk, where the temperature gradient is very small and work our way to the midplane. At each location the criterion according to Eq.~\ref{eq:convection} is determined. If the temperature gradient is too steep the temperature gradient is adjusted such that Eq.~\ref{eq:convection} becomes an equality. In some cases this causes the temperature structure close to midplane to be adjusted such that the midplane temperature is reduced. This way we can model the effects of cooling of the midplane by convection although we should stay aware of the approximate nature of the implementation.

Our implementation of convection is very similar to the one by \citet{1998ApJ...500..411D} with the difference that we do not consider the turbulent flux, and we take the convection efficiency, their parameter $\zeta$, to be unity. In the relatively less dense regions of the disk transfer of energy through radiation dominates and the criterion Eq.~(\ref{eq:convection}) is fulfilled. In this case we do not alter the temperature structure. However, in those regions of the disk where radiative transfer cannot fulfill Eq.~(\ref{eq:convection}) we do adjust the temperature structure to be adiabatic. This means that convection becomes important only in the more massive disks where radiative energy transfer is difficult.

\section{Results}

\begin{figure}[!tb]
\centerline{\resizebox{0.9\hsize}{!}{\includegraphics{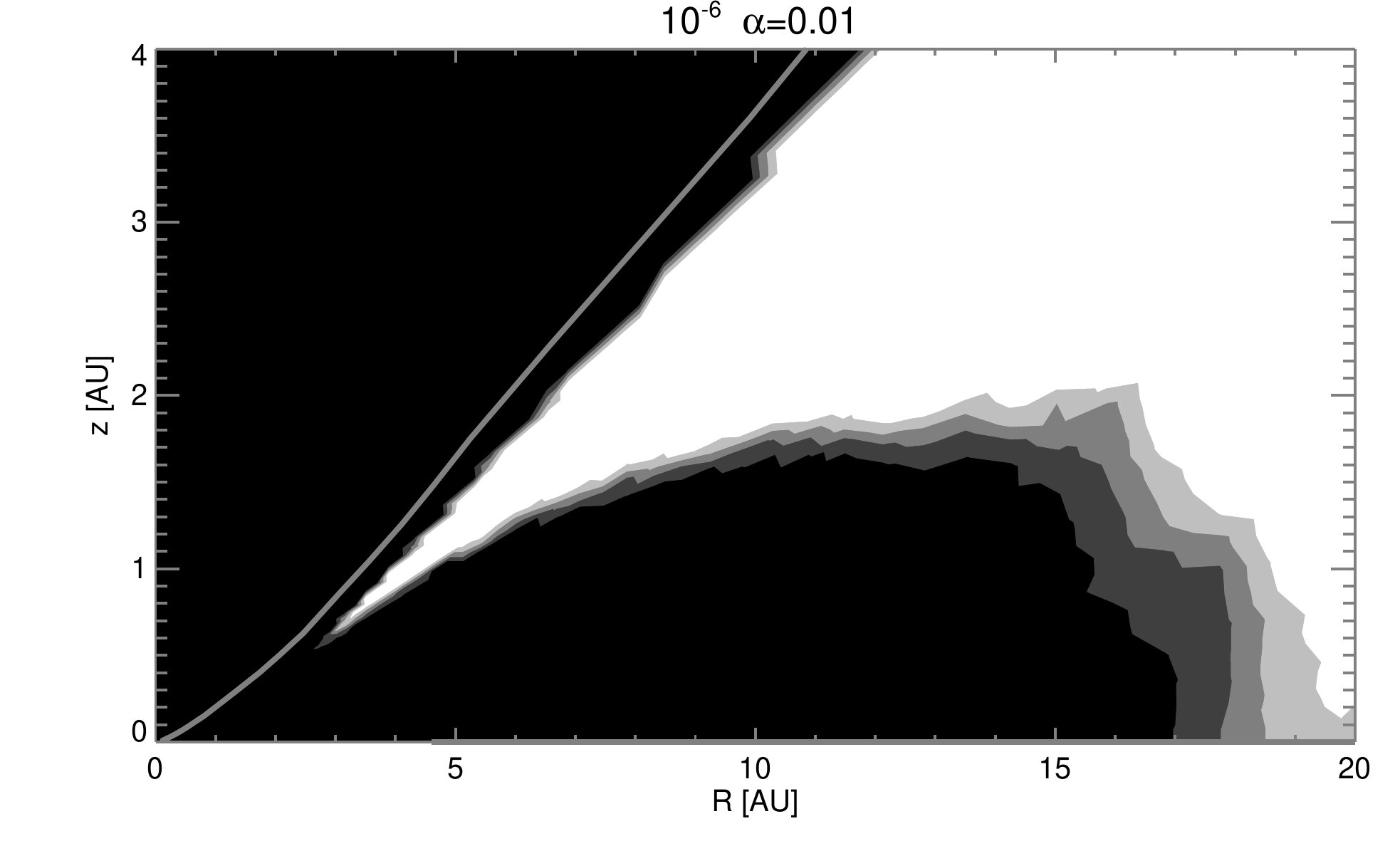}\includegraphics{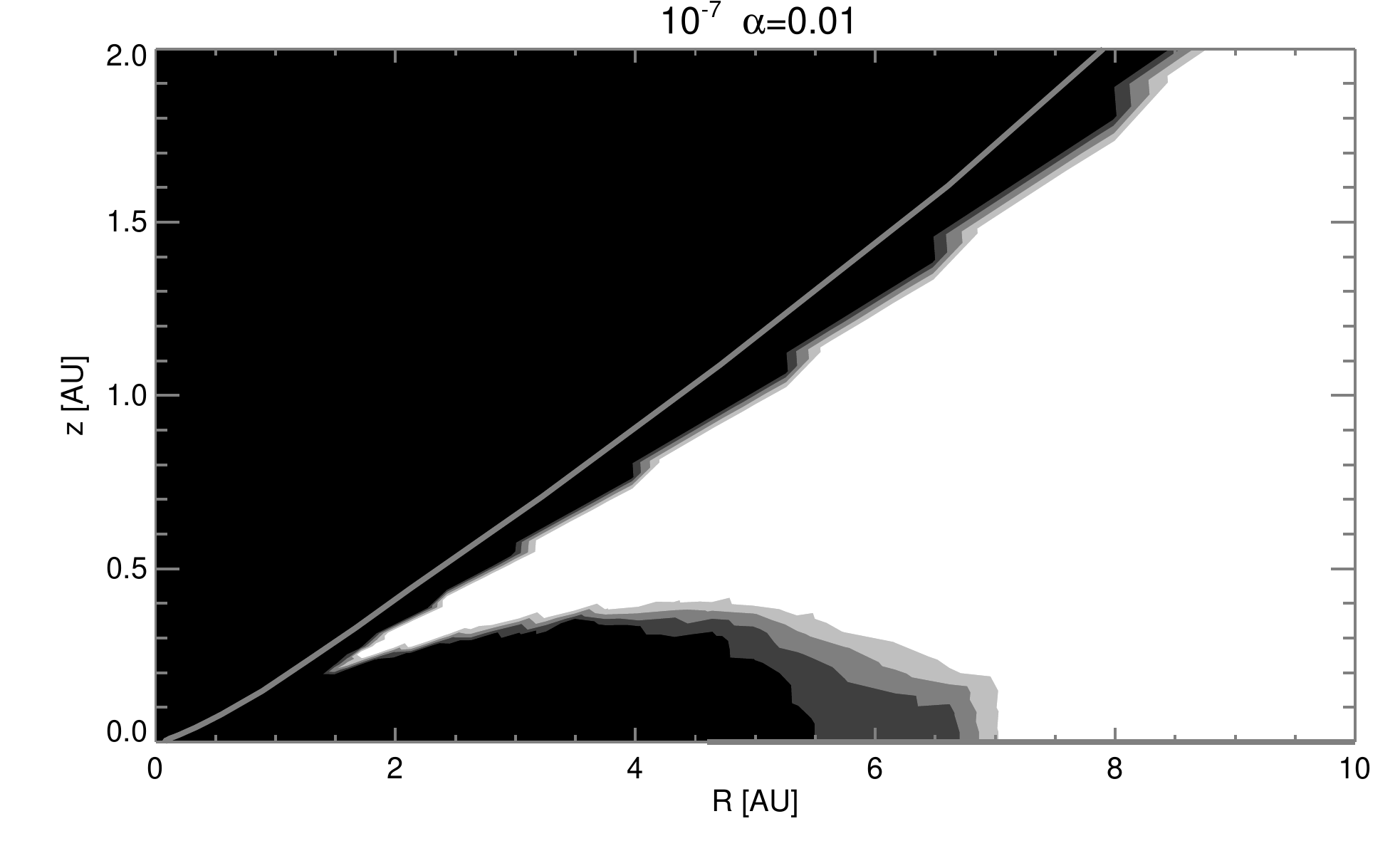}}}
\centerline{\resizebox{0.9\hsize}{!}{\includegraphics{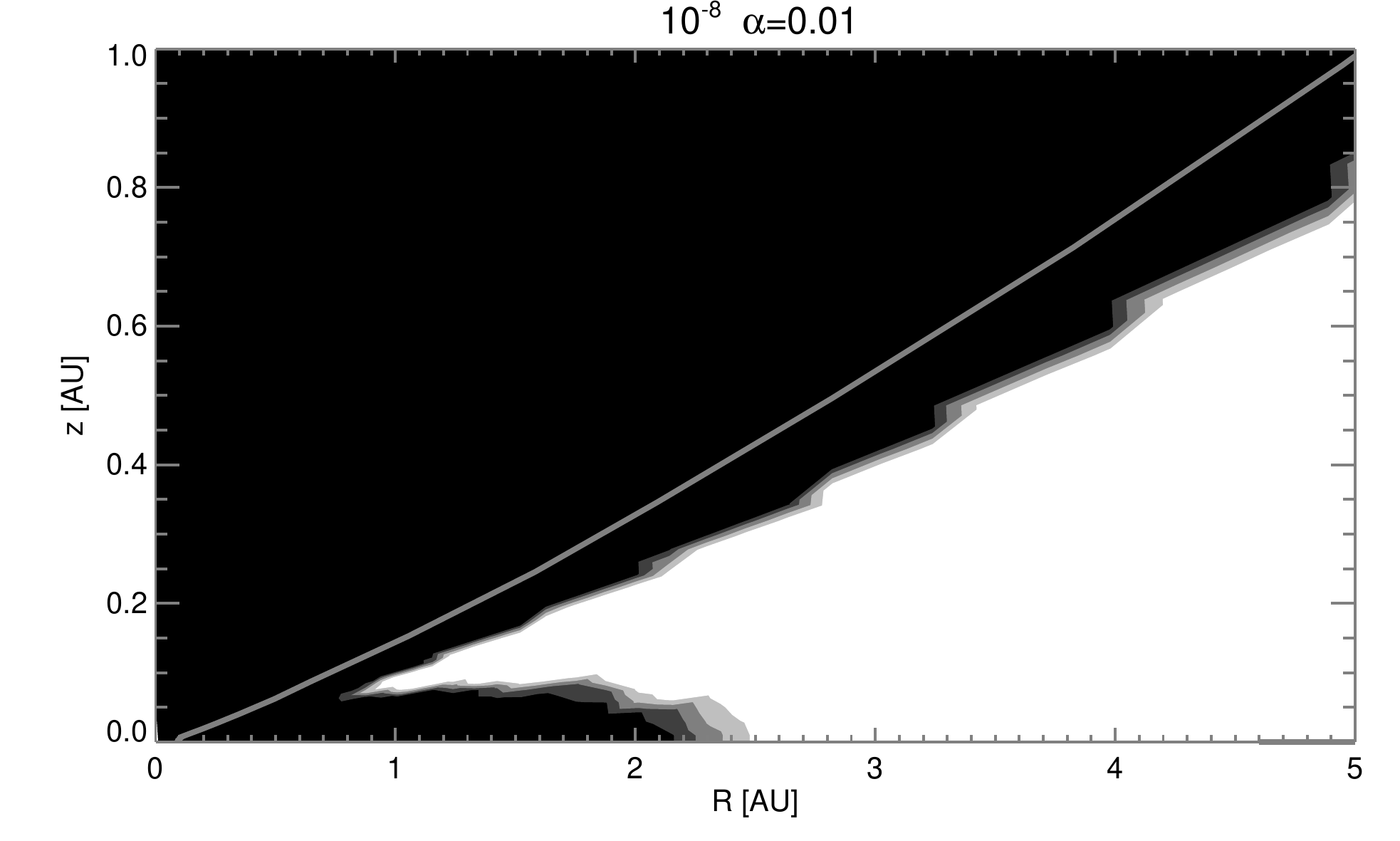}\includegraphics{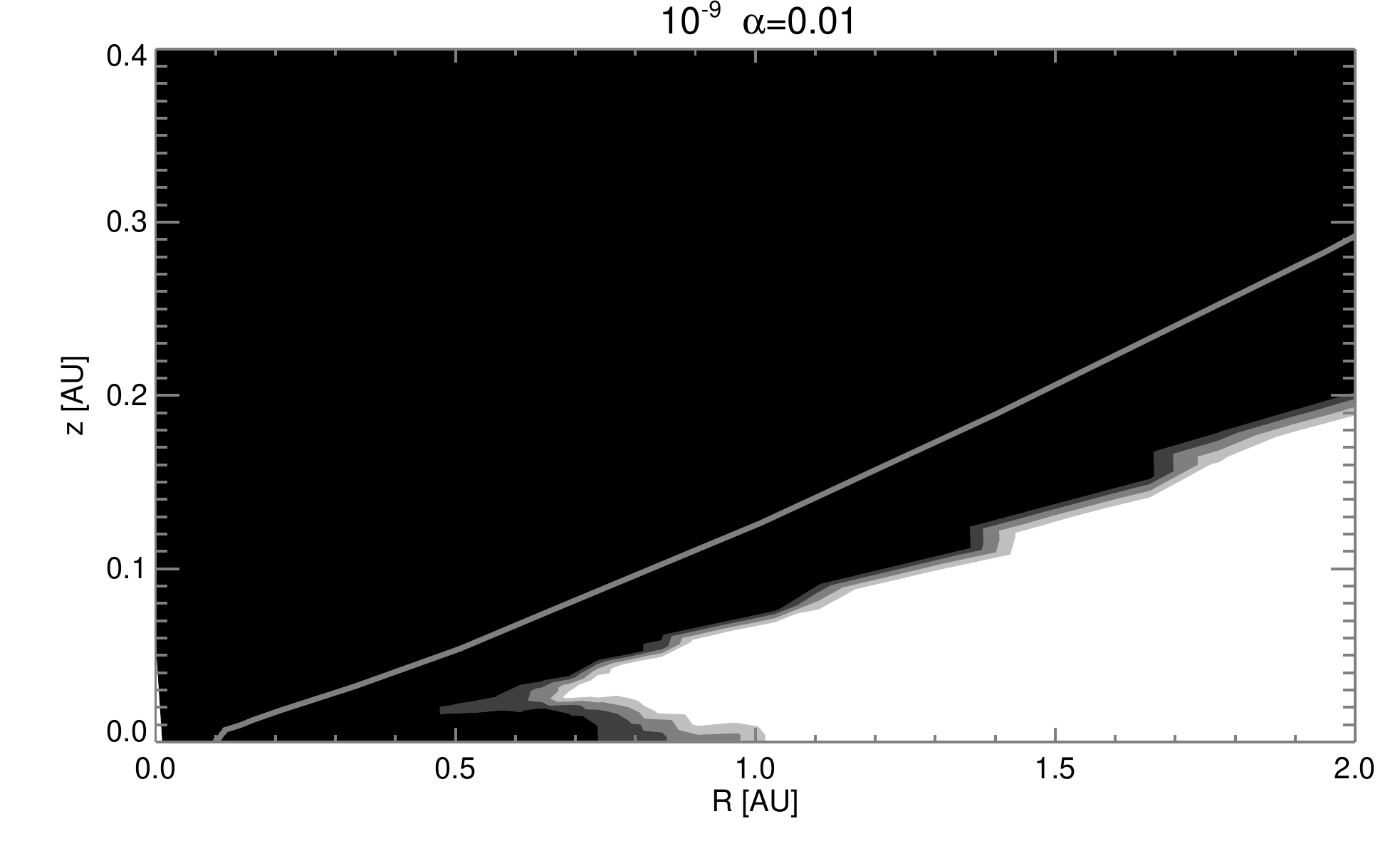}}}
\caption{Same as Fig.~\ref{fig:snow lineCarbon0.5a0.01} but for the case where all carbon is in the form of solid carbon.} 
\label{fig:snow lineCarbon0.01}
\end{figure}

\subsection{Location of the snow line as a function of $\dot M$}

In Fig.~\ref{fig:snow lineCarbon0.5a0.01} we display the locations in the disk where water ice can exist for various values of the mass accretion rate. It is clear that the location of the snow-line depends strongly on the value of $\dot{M}$. This is caused by two main effects. The first is that when $\dot{M}$ is increased, the energy released by viscous stress is increased. Secondly, a larger value of $\dot{M}$ causes the disk to be more massive (see also Fig.~\ref{fig:surfacedens}) and thus the greenhouse effect, keeping the energy at the midplane, is much stronger (as explained below, see also Eq.~(\ref{eq:midplaneT1})) and the midplane temperature is increased further.

Going vertically up in the disk, the effect from viscous heating becomes less important and the snow line moves in significantly. This continues up to the point where radiation from the Sun becomes the dominant energy source and causes the snow line to move out again. The fact that the location of the snow line in the upper, optically thin region of the disk is not constant is caused by the pressure dependent evaporation temperature of the ice in combination with a decreasing density when going higher up in the disk. This shape of the ice region is consistent with the findings from \citet{2005ApJ...620..994D} but the snow line in our computations lies significantly further out. The reason for this is most likely predominantly caused by the different opacity of the grains which is assumed. \citet{2005ApJ...620..994D} use for the opacity the analytical expression from \citet{1986ApJ...308..883R}. As outlined below the most important parameter determining the midplane temperature is the Rosseland mean opacity at 160\,K divided by the gas to solid ratio, since this determines how efficiently the energy created by viscous stress is kept in the midplane at the location of the ice evaporation front. In the study by \citet{2005ApJ...620..994D} this parameter is $\kappa_R/f=2.7\,$cm$^2$/g, while in our study, taking a realistic dust grain model, we find that $\kappa_R/f$ lies in the range $5.9$ to $11.1\,$cm$^2$/g (see Table~\ref{tab:abundances}). Thus in our model the energy generated by viscous heating is more efficiently captured in the midplane regions.

\begin{figure}[!tb]
\centerline{\resizebox{0.9\hsize}{!}{\includegraphics{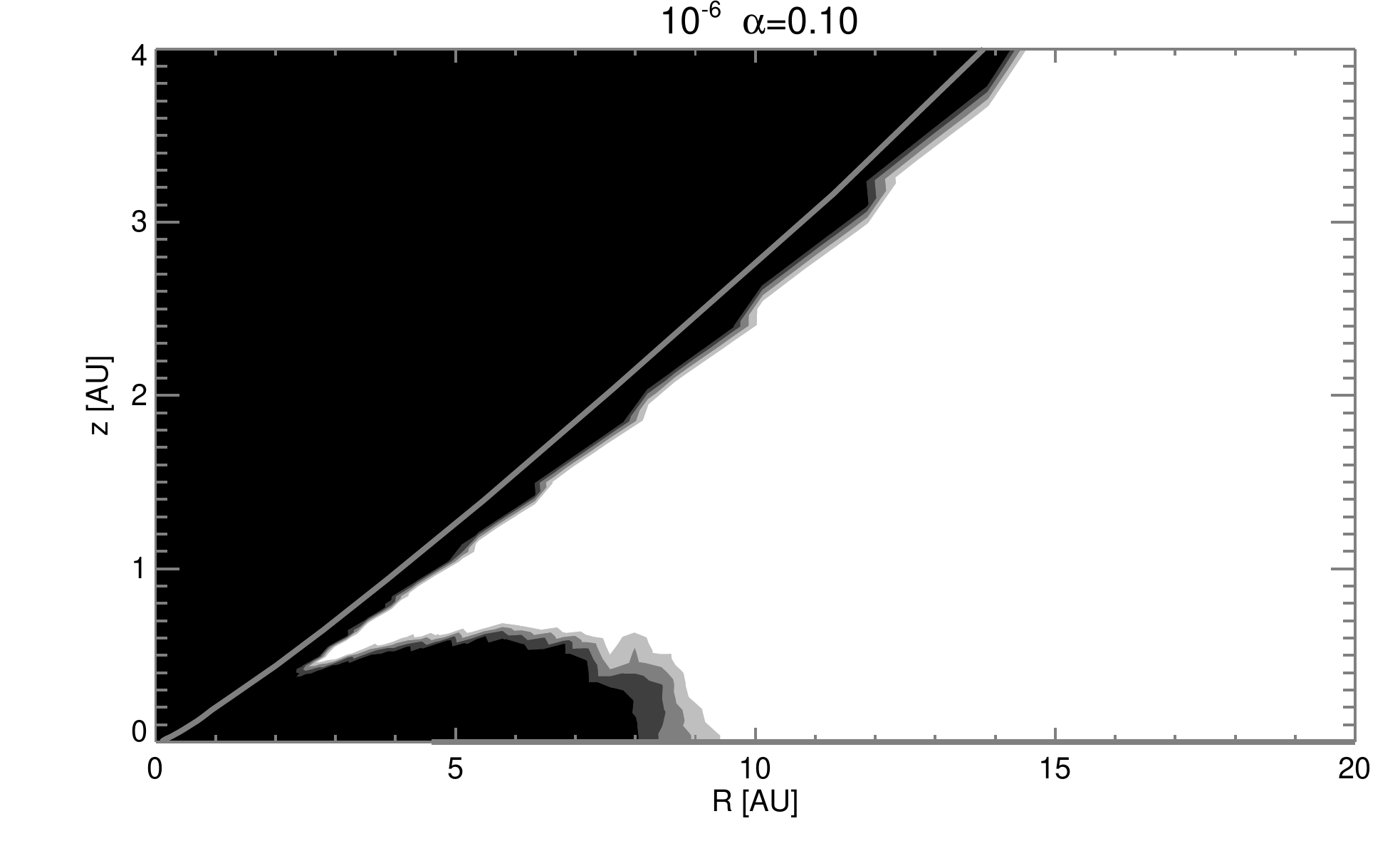}\includegraphics{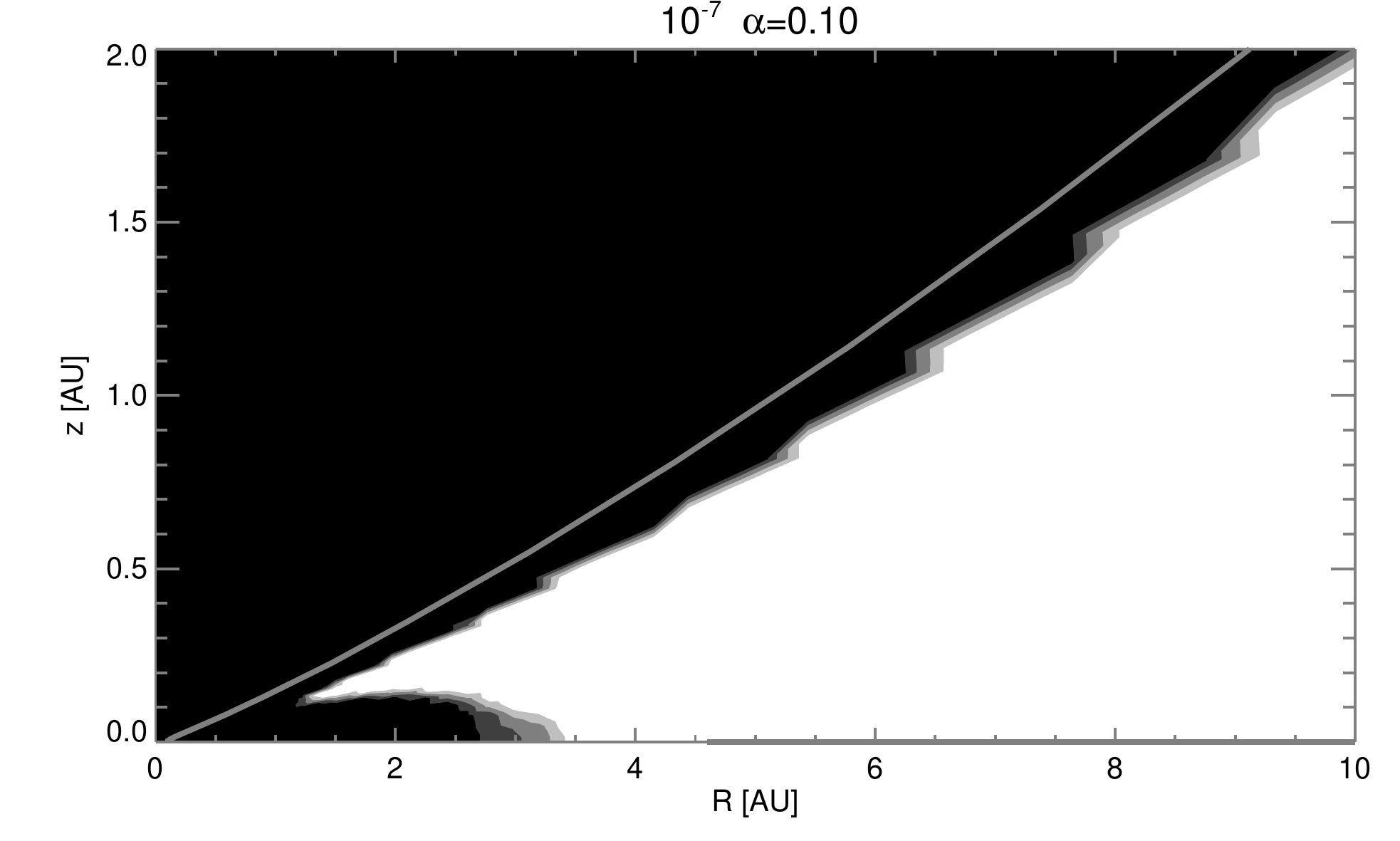}}}
\centerline{\resizebox{0.9\hsize}{!}{\includegraphics{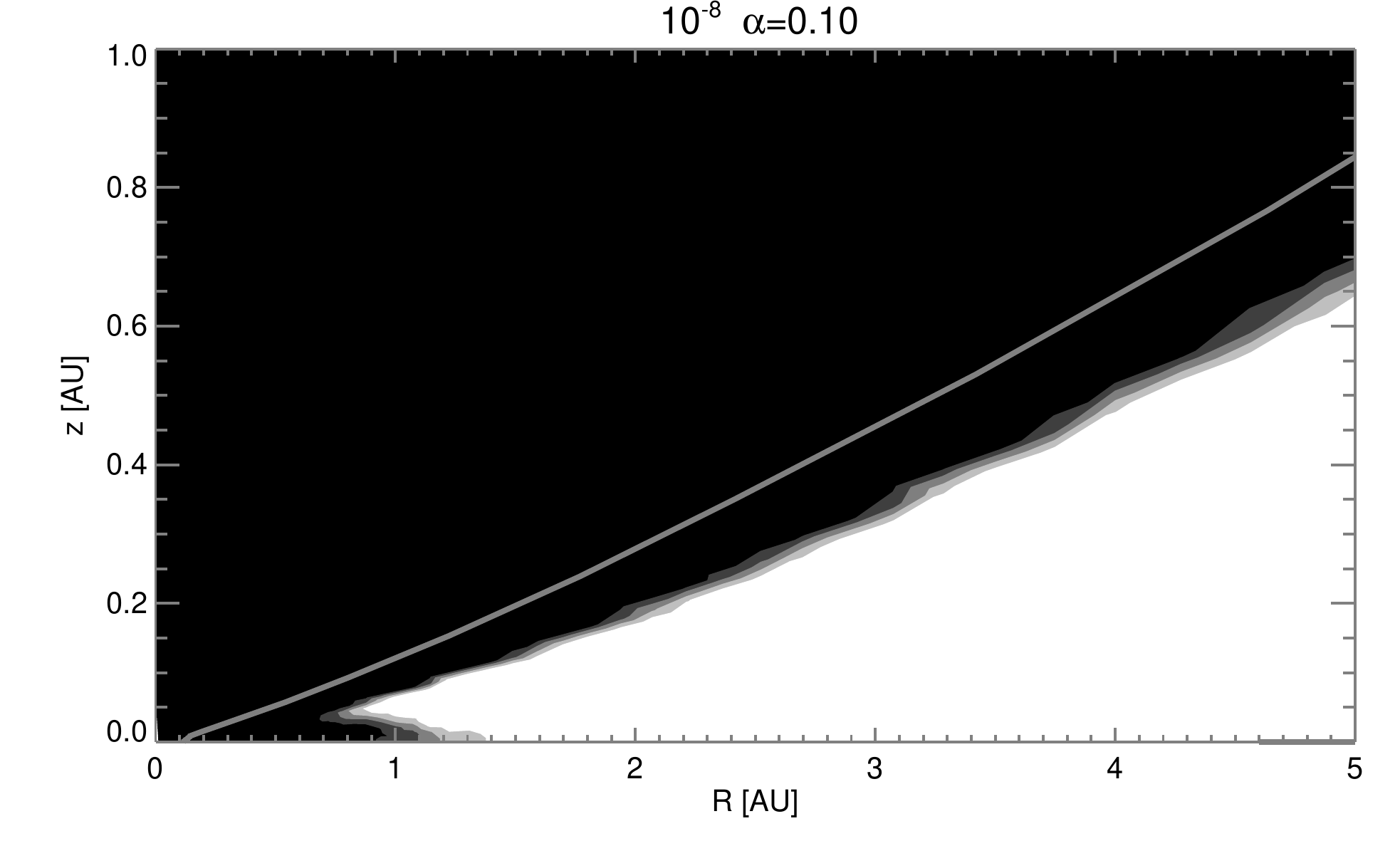}\includegraphics{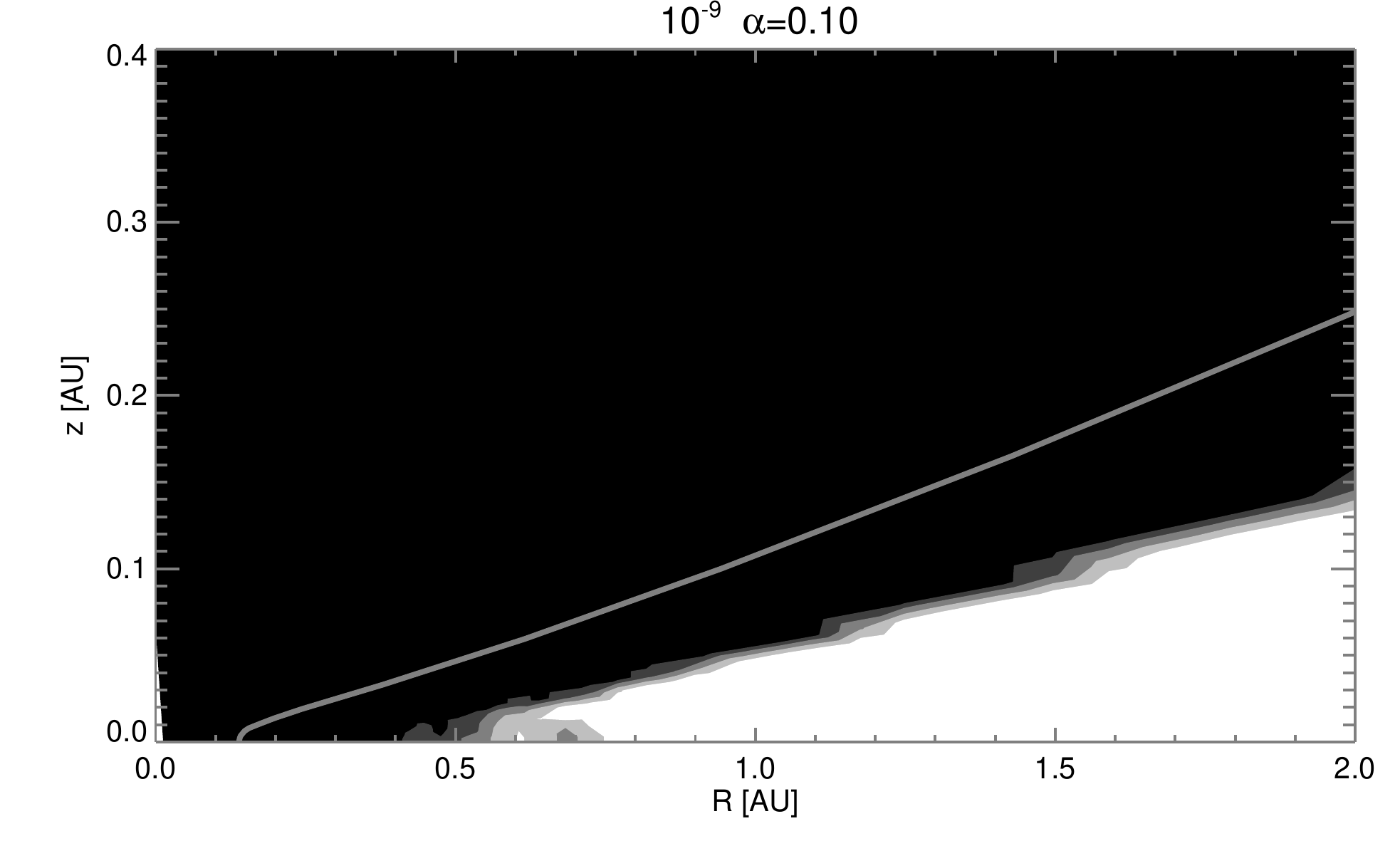}}}
\caption{Same as Fig.~\ref{fig:snow lineCarbon0.5a0.01} but for the case where all C atoms are in the form of CO and $\alpha=0.1$.} 
\label{fig:snow lineCO0.10}
\end{figure}

In Figs.~\ref{fig:snow lineCarbon0.01} and \ref{fig:snow lineCO0.10} we show the extreme cases of the range of dust parameters we studied. Fig.~\ref{fig:snow lineCarbon0.01} shows the case where all carbon atoms are locked into solid carbon, i.e. the opacity is at its maximum. In this case the greenhouse effect, heating the midplane efficiently is at its maximum. The other extreme, presented in Fig.~\ref{fig:snow lineCO0.10}, shows the case where all carbon is in CO, i.e. the opacity is minimum, and the turbulence is very high, $\alpha=0.1$. 
This high value of $\alpha$ implies that one can get the same mass accretion rate with a smaller surface density. Thus for a given value of $\dot{M}$ the surface density drops and the greenhouse effect is smaller. In this case the energy can easily escape the midplane and the midplane temperature is low leading to a snow line close to the star.

\subsection{Analytical estimate for the location of the snow line in the midplane}

We take the analytic estimate of the midplane temperature from \citet{1990ApJ...351..632H}. This only includes viscous heating in the midplane and results in a midplane temperature, assuming high optical depths, of
\begin{equation}
\label{eq:midplaneT1}
T^4=\frac{9\Sigma_\mathrm{gas} \kappa_R G M_\star \dot{M}}{128\pi \sigma R^3 f},
\end{equation}
where $\kappa_R$ is the Rosseland mean opacity of the gas/ice mixture and $f$ is the gas to solid ratio such that the total optical depth through the disk is equal to $\tau_\mathrm{tot}=\kappa_R\Sigma_\mathrm{gas}/f=\kappa_R\Sigma_\mathrm{solid}$. The values of $\kappa_R$ and $f$ for the models we computed are listed in Table~\ref{tab:abundances}. The surface density of the gas, $\Sigma_\mathrm{gas}$, can be expressed in terms of $T$ and $\alpha$ using Eqs.~(\ref{eq:gamma}) and (\ref{eq:tot energy}). The integral of Eq.~(\ref{eq:gamma}) in vertical direction is equal to the total energy given in Eq.~(\ref{eq:tot energy}). Assuming $R>>R_\star$ and taking the temperature to be the midplane temperature at each height in the disk we get \citep[see also][]{2007prpl.conf..555D}
\begin{equation}
\label{eq:surfdens}
\Sigma_\mathrm{gas}=\frac{\mu m_p \dot{M}}{3\pi \alpha k_b T}\sqrt{\frac{G M_\star}{R^3}},
\end{equation}
where $\mu$ is the average molecular mass and $m_p$ is the proton mass. Substituting Eq.~(\ref{eq:surfdens}) into Eq.~(\ref{eq:midplaneT1}) we get
\begin{equation}
\label{eq:midplaneT2}
T^5=\frac{3 \mu m_p \dot{M}^2 \kappa_R}{128\pi^2 \alpha k_b \sigma f} \left[\frac{G M_\star}{R^3}\right]^\frac{3}{2} .
\end{equation}

Taking the above we find that the snow line is located at
\begin{equation}
\label{eq:Restimate}
R_\mathrm{ice}=\left[C\cdot
\frac{
\dot{M}^2
\kappa_R}{f \alpha}
\right]^\frac{2}{9},
\end{equation}
where $\kappa_R$ is the Rosseland mean opacity of the gas/ice mixture computed at 160\,K (see Table~\ref{tab:abundances}), and $C$ is given by
\begin{equation}
C=\frac{3\mu m_p \left(G M_\star\right)^{3/2}}{128\pi^2k_b\sigma T_\mathrm{ice}^5}=1.7\cdot10^{22}\,\mathrm{cm^{5/2}\,s^2/g}.
\end{equation}
We find that this formula agrees very well with our findings for the range of parameters considered here (within a few percent). The formula becomes less accurate when the snow line is close to the star, and clearly the equation fails when the radiation of the Sun is the dominant energy source in the midplane. However, this is never the case in the models considered here. For easy computation, the value of $C=3.5\cdot10^{14}$ can be used in Eq.~\ref{eq:Restimate} when $\dot{M}$ is given in $M_{\sun}/$yr and $\kappa_R$ is given in cm$^2/$g to give the location of the snow line $R_\mathrm{ice}$ in AU.

Using Eq.~\ref{eq:Restimate} provides a way to scale our results to those obtained by \citet{2005ApJ...620..994D} taking into account the correct values of $\kappa_R$, $f$, and $\dot{M}$. Doing this we find that the agreement is excellent, and thus the differences are indeed due to the different values of $\kappa_R/f$. We also compared our results to those computed by \citet{2007ApJ...654..606G}. The values for the location of the snowline as computed using Eq.~(\ref{eq:Restimate}) are compared to those obtained by \citet{2005ApJ...620..994D} and \citet{2007ApJ...654..606G} in table \ref{tab:compare}. Note that \citet{2005ApJ...620..994D} adopted a different description for the viscosity of the disk, which explains partly the differences. The overall agreement with the approximated and numerically computed values implies that estimates of the location of the snow line using approximate methods are in general correct when appropriate opacity values are used.

\begin{table*}[!tbp]
\begin{center}
\linespread{1.3}%
\selectfont
\begin{tabular}{lccc}
$\dot{M}$ [$M_\sun/$yr]	&	$\kappa_R/(f\alpha)$ [cm$^2$/g]	&	$R_\mathrm{ice}$ [AU] (Eq.~\ref{eq:Restimate})	&	$R_\mathrm{ice}$ [AU] (numerical) \\
\hline
\multicolumn{4}{l}{\citep{2005ApJ...620..994D}}\\
$8.06\cdot10^{-6}$	&	$\sim270$		&	32.3		&	$\sim37$\\
$1.75\cdot10^{-6}$	&	$\sim270$		&	16.4		&	$\sim15.5$\\
$1.13\cdot10^{-7}$	&	$\sim270$		&	4.8		&	$\sim3.6$\\
$2.53\cdot10^{-10}$	&	$\sim270$		&	0.3		&	$\sim0.6$\\
\hline
\multicolumn{4}{l}{\citep{2007ApJ...654..606G}}\\
$1\cdot10^{-7}$	&	$27.7$		&	2.8		&	$\sim3.4$\\
$1\cdot10^{-8}$	&	$27.7$		&	1.0		&	$\sim1.4$\\
$1\cdot10^{-9}$	&	$27.7$		&	0.4		&	$\sim0.6$\\
\hline
\multicolumn{4}{l}{This study ($w=0.5, \alpha=0.01$)}\\
$1\cdot10^{-6}$		&	770	&	16.1		&	$\sim16$\\
$1\cdot10^{-7}$		&	770	&	5.8		&	$\sim6$\\
$1\cdot10^{-8}$		&	770	&	2.1		&	$\sim2.2$\\
$1\cdot10^{-9}$		&	770	&	0.7		&	$\sim0.8$\\
\hline
\multicolumn{4}{l}{This study ($w=0, \alpha=0.1$)}\\
$1\cdot10^{-6}$		&	59	&	9.1		&	$\sim9$\\
$1\cdot10^{-7}$		&	59	&	3.3		&	$\sim3.2$\\
$1\cdot10^{-8}$		&	59	&	1.2		&	$\sim1.2$\\
$1\cdot10^{-9}$		&	59	&	0.4		&	$\sim0.5$\\
\end{tabular}
\linespread{1}%
\selectfont
\end{center}
\caption{A comparison of the location of the snow line as obtained using Eq.~(\ref{eq:Restimate}) and using numerical computations. The values from the studies by \citet{2005ApJ...620..994D} and \citet{2007ApJ...654..606G} are read from the figures. The value of $\kappa_R/(f\alpha)$ for the study by \citet{2005ApJ...620..994D} is an estimate since they use a different description for the viscosity which corresponds approximately to ours with $\alpha=0.01$. For the study by \citet{2007ApJ...654..606G} we used their Eq.~(2) to determine the value of $\kappa_R$ at 160\,K. The values of $R_\mathrm{ice}$ for the numerical computations of this study correspond to the location where 50\% of the ice is condensed.}
\label{tab:compare}
\end{table*}

\subsection{Solid surface density distribution as a function of $\dot M$}

\begin{figure}[!tb]
\centerline{\resizebox{0.7\hsize}{!}{\includegraphics{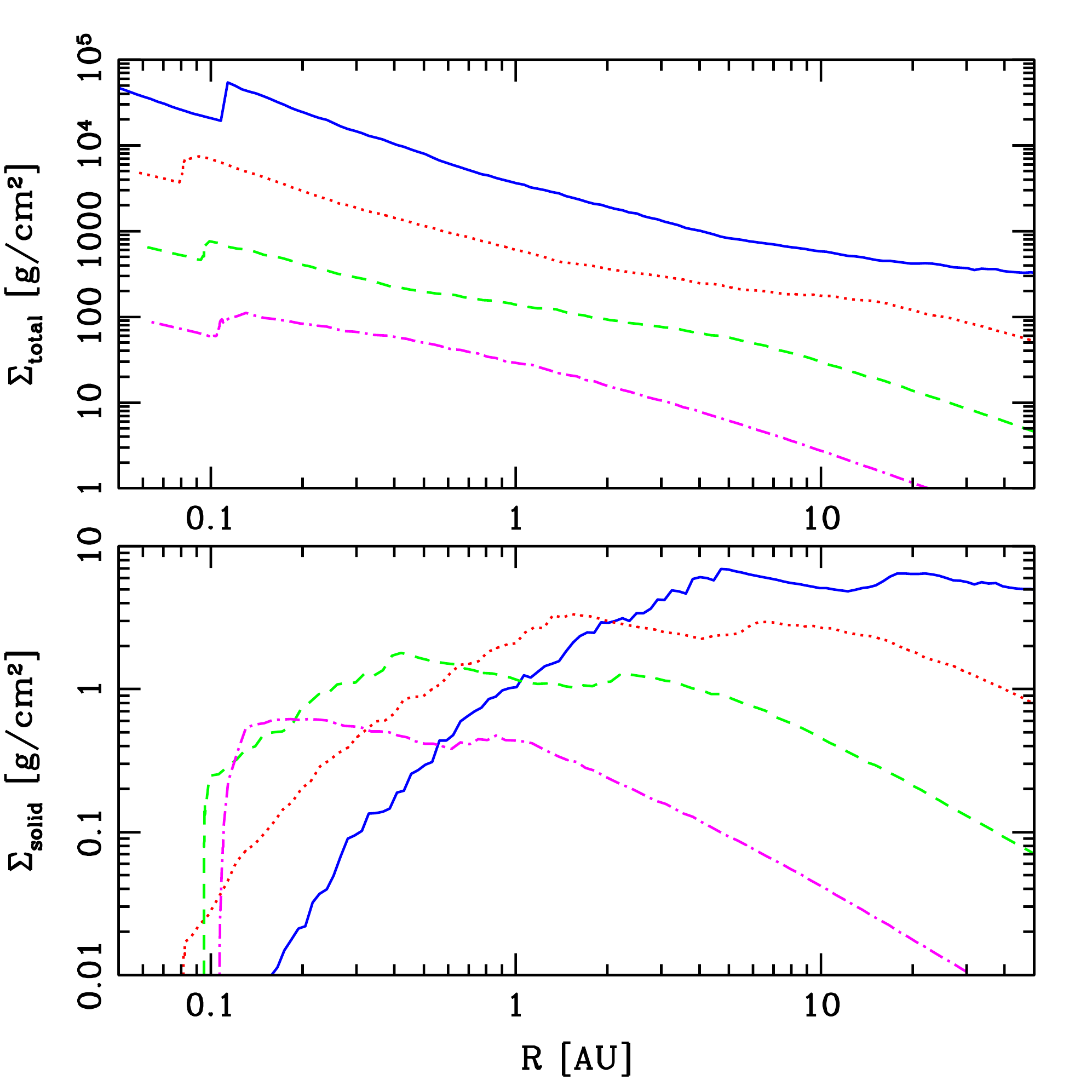}}}
\caption{The surface density as a function of radius for different values of the mass accretion rate. The top panel shows the total surface density, while the lower panel shows the surface density in solids. Blue solid lines are for $\dot{M}=10^{-6}\,M_\sun/$yr, red dotted lines for $\dot{M}=10^{-7}\,M_\sun/$yr, green dashed lines are for $\dot{M}=10^{-8}\,M_\sun/$yr, and purple dashed-dotted lines are for $\dot{M}=10^{-9}\,M_\sun/$yr. In all cases half of the available carbon atoms are in the solid phase and $\alpha=0.01$. } 
\label{fig:surfacedens}
\end{figure}

In Fig.~\ref{fig:surfacedens} we show the surface density as a function of radius for different values of the mass accretion rate. Both the total surface density (top panel) as well as the surface density in solids (lower panel) are shown. The sudden drop in the total surface density at the location where the solids start to become important is an effect of the sudden increase in opacity, causing a steep temperature gradient which translates into a steep surface density gradient through Eqs.~\ref{eq:gamma} and \ref{eq:tot energy}. The surface density in solids is adjusted iteratively until the temperature nowhere in the disk exceeds the sublimation temperature. This means in the inner regions solid matter is removed until the greenhouse effect, keeping the energy locked in the midplane regions of the disk, drops to a level where the temperature is low enough. For very high accretion rates, a lot of matter has to be removed to let the large amount of energy produced in the midplane escape. This balance between condensation and evaporation results in a thermostat keeping the midplane temperature at the silicate evaporation temperature over a significant part of the inner disk. When the accretion rate is reduced, more matter can condense. However, at the same time reducing the mass accretion rate reduces the total surface density and thus the amount of matter that can condense. Both these effects play a role in determining the surface density in the solid phase.

At the radius of water ice sublimation we have a gradual increase in the surface density in the solid phase. We find that the increase in the solid phase due to water ice is $F_\mathrm{ice}=1.84$ for the case where half of the carbon atoms is in the solid phase. This increase is quite small compared to the $F_\mathrm{ice}=4.2$ taken currently in the literature. We have to take into account here that we have only considered water ice condensation. If we take in addition CH$_4$ and NH$_3$ ice, which are the most important ice species after water ice, we gain about a factor 1.5 \citep[see e.g.][]{2009Icar..200..672D} so we get $F_\mathrm{ice}\sim2.8$. Note however, that these ices condense at lower temperatures, so somewhat further out then the water-ice condensation radius we compute here.

The surface density is in principle a function of the mass accretion rate, the value of $\alpha$ and the opacity of the grains. However, in the innermost region the dust evaporates to reduce the greenhouse effect such that the midplane temperature drops below the dust sublimation temperature. In this regime the solid surface density is adjusted to reach a given vertical optical depth, $\kappa_R\Sigma_\mathrm{gas}/f$, roughly following Eq.~(\ref{eq:midplaneT1}). The maximum vertical optical depth that can exist before the midplane gets too hot is thus only a function of the total energy released in the midplane, and thereby only of $\dot{M}$ (see Eq.~\ref{eq:tot energy}). The amount of mass that can exist in this region is robustly determined through the mass accretion rate and the opacity of the grains and is independent of the gas density or the description of the accretion mechanism. Therefore, in this region we can easily scale the solid surface density as a function of $\dot{M}$ we find in this study to different values of the opacity of the grains when, for example, they grow to larger sizes.

\subsection{Effects of convection}

When studying the temperature structure in the disk in detail we find that often the temperature gradient is steeper than the adiabatic temperature gradient (see Eq.\ref{eq:convection}). In these cases the midplane will cool by convection. If we correct for the convective cooling in the way discussed in section \ref{sec:convection}, the midplane temperature decreases in some cases significantly. We find that this only plays a role when the midplane temperature is dominated by accretion and in addition, is most important when the temperature is above $\sim500\,$K. Thus, the location of the snow line is only mildly affected. For the case of $\dot{M}=10^{-6}\,M_{\sun}/$yr we find that due to the decreased midplane temperature the fraction of condensed solids can increase locally by a factor of 5 with respect to the case where convection is ignored. For low accretion rates convection becomes unimportant at almost all radii and the disk cools through radiation \citep[see also][]{1998ApJ...500..411D}. The reason for the dependence of the importance of convection on the surface density of the disk is simply due to the fact that less massive disks can easily cool through radiation, so convection is not needed to lose the energy from the midplane. In the more massive disks it is very difficult for the radiation to escape from the midplane and convective cooling is much more efficient. Our treatment of convection is somewhat simplified and more advanced methods, also including turbulent fluxes, should be implemented in the future to get a clear picture of the exact effects.

\begin{figure}[!tb]
\centerline{\resizebox{0.55\hsize}{!}{\includegraphics{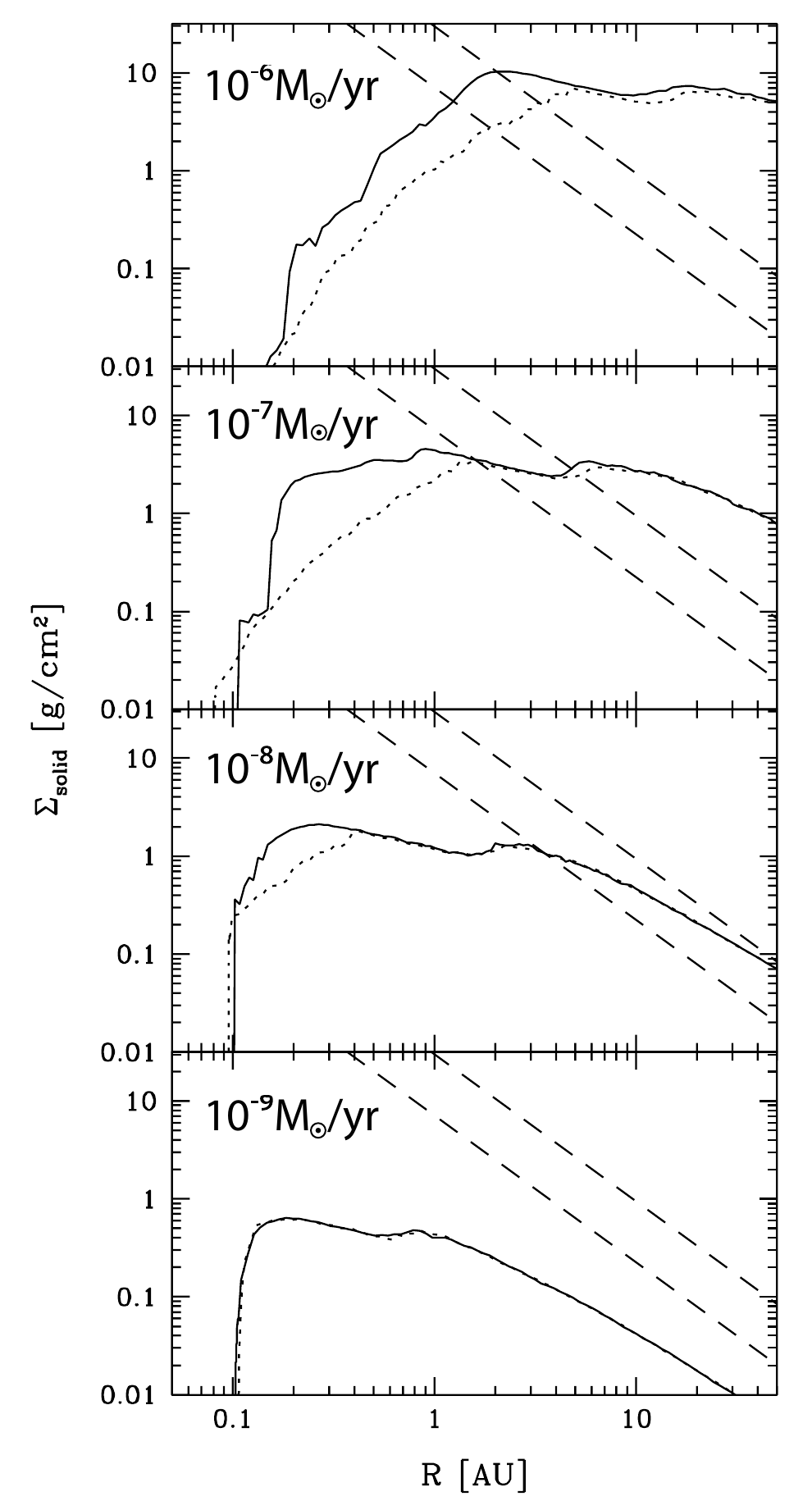}}}
\caption{The surface density as a function of radius for different values of the mass accretion rate. The solid lines are with convective cooling included, the dotted lines are when only radiative cooling is used (same as in Fig.~\ref{fig:surfacedens}). In all cases half of the available carbon atoms can be in the solid phase and $\alpha=0.01$. In all panels the dashed lines gives the MMSN according to Eq.~\ref{eq:mmsn} with and without ice.}
\label{fig:convection}
\end{figure}

In Fig.~\ref{fig:convection} we plot the surface density in the solid phase for computations with and without convective cooling of the midplane. It is clear that for the high accretion rates convective cooling is important to gain extra solids in the region where the terrestrial planets are (i.e. in the inner few AU). When convective cooling is ignored we find that there is never enough mass in the solids around 1\,AU to create the Earth. However, with convective cooling we find that there is much more mass. Note that the amount of mass in the solids can be increased even further for the high accretion phases by growing the grains to somewhat larger sizes, thereby decreasing the opacity and thus the midplane temperature. We conclude that convective cooling is an important mechanism to take into account when modeling the solid density distribution in the early Solar nebula.

\section{Discussion}

\subsection{Implications for planet formation}

In this section we compare our results with constraints from the minimum mass solar nebula (MMSN). We will use here the 'classical' MMSN as mentioned above. There is still quite some debate on what the minimum mass is needed to form our solar system \citep[see e.g.][]{2005ApJ...627L.153D, 2007ApJ...671..878D, 2009ApJ...698..606C}, and most of these models find the need for a more massive solar nebula than the classical one we use here. Therefore, our choice of the MMSN is conservative, and the computations could be repeated to find stronger constraints by using other models. In addition, planet formation might be more dynamic, i.e. planets could accrete mass from different parts of the disk through dynamic interactions.

According to the MMSN (Eq.~\ref{eq:mmsn}) in order to form the Earth one needs a solid surface density of 7.1\,g/cm$^2$ at 1\,AU. As can be seen directly from Fig.~\ref{fig:convection} this surface density in solids is never achieved in our standard model. This is because for the high accretion models the midplane gets too hot, so solids have to evaporate to let the energy escape, while for the low accretion models there simply is not enough mass at all. However, as discussed above, if we take into account convective cooling of the midplane the amount of condensed solids can increase by a factor of 5. This way the surface density at 1\,AU is only slightly below the requirement from the MMSN. Inside 1\,AU the surface density in solids is never even close to the MMSN, which might pose serious problems for the formation scenario of Venus.

Jupiter is generally assumed to form outside the snow line. Thus, when Jupiter formed at its current position, we have as an additional constraint that $R_\mathrm{ice}<5.2\,$AU. For this the mass accretion rate needs to be smaller than $\dot{M}<8\cdot10^{-8}\,M_{\sun}/$yr. Since the midplane temperature at the snow line should be smaller than 160\,K convective cooling is not important, so we can use our standard model. At this mass accretion rate the solid surface density is approximately $\Sigma_\mathrm{solid}\approx 2\,$g/cm$^2$, which is slightly lower than the mass needed according to Eq.~\ref{eq:mmsn}. However, typical giant planet formation models require up to 3 to 4 times the value of the MMSN \citep[see e.g][]{1996Icar..124...62P}.

We thus find that our computations put constraints on the scenario of the formation of the Solar system in the context of the MMSN. However, as we already noted above, the MMSN might be an optimistic estimate, and in reality a much larger mass could be required. A possible solution to still meet the requirements of planet formation in that case would be to lower the opacity of the grains significantly by, for example, growing them to larger sizes.

\subsection{Dust properties in the hot midplane regions}

We find that at high mass accretion rates there is a significant region in the disk where the midplane temperature is above 1000\,K. This is out to $\sim$1.7\,AU in the case where $\dot{M}=10^{-6}\,M_{\sun}$/yr, and out to $\sim$0.6\,AU in the case of $\dot{M}=10^{-7}\,M_{\sun}$/yr. This is including convective cooling. Thus we find that a considerable region in the early Solar system had conditions where thermal processing of the solid state material was very efficient.
Also, at these temperatures grain coagulation is influenced by \emph{sintering} \citep[see e.g.][]{2003Icar..164..139P}. Sintering is the process where the bonds between grains are molten together by the impact of grain collisions or by a long exposure to high temperatures. The structures formed under these conditions can be extremely open aggregates when the strong bonds created do not allow restructuring or compaction of the aggregates. \citet{2003Icar..164..139P} compute that for aggregates composed of silica spheres exposure to temperatures around $\sim1250$\,K is needed for efficient sintering. This is comparable to the crystallization temperature of silica found by \cite{2000A&A...364..282F}. Since both sintering and crystallization are processes that are caused by the atoms in the lattice wanting to move to the minimum energy state, it is not unlikely that the temperatures of these processes are similar. In that case sintering of Fe/Mg silicates will happen at slightly lower temperatures, i.e. around $\sim1000$\,K. The collision speeds in the midplane regions of the disk are too low to cause collisional sintering.

As mentioned before, grain growth might help to increasing the solid surface density by reducing the opacity of the particles and in this way decreasing the temperature in the disk. A low opacity reduces the greenhouse effect and thus allows more grains to condense before the midplane temperature gets too high. In fact, when the Rosseland mean opacity computed at the dust evaporation temperature of 1500\,K reduces by a given factor, the maximum allowed surface density increases by this factor. In order to test to what sizes we have to grow the grains in order to significantly increase the solid surface density we performed computations of the Rosseland mean opacity at 1500\,K for larger grain sizes.
We increased the entire size distribution by a given factor $\gamma$ which means that we do not only grow large particles, but efficiently remove the small particles as well. Thus, where the MRN distribution we used so far runs from $0.005\,\mu$m to $0.25\,\mu$m, the new size distribution contains grains with sizes from $(\gamma\times0.005)\,\mu$m to $(\gamma\times0.25)\,\mu$m.
For these larger grain sizes the effects of anisotropic scattering become significant. The argumentation on the maximum total optical depth were made under the assumption of isotropic scattering. When anisotropic scattering is important, the same argumentation can be made when scaling the scattering cross section by a factor $(1-g)$ \citep[see][]{1978usaf.book.....I} where $0<g<1$ is the anisotropy parameter. For values of $g$ close to unity, most radiation is scattered in forward direction which for radiative transfer purposes is effectively a strong reduction of the effect of scattering. Values of $g$ close to zero are more equivalent to isotropic scattering.

\begin{figure}[!tb]
\centerline{\resizebox{0.7\hsize}{!}{\includegraphics{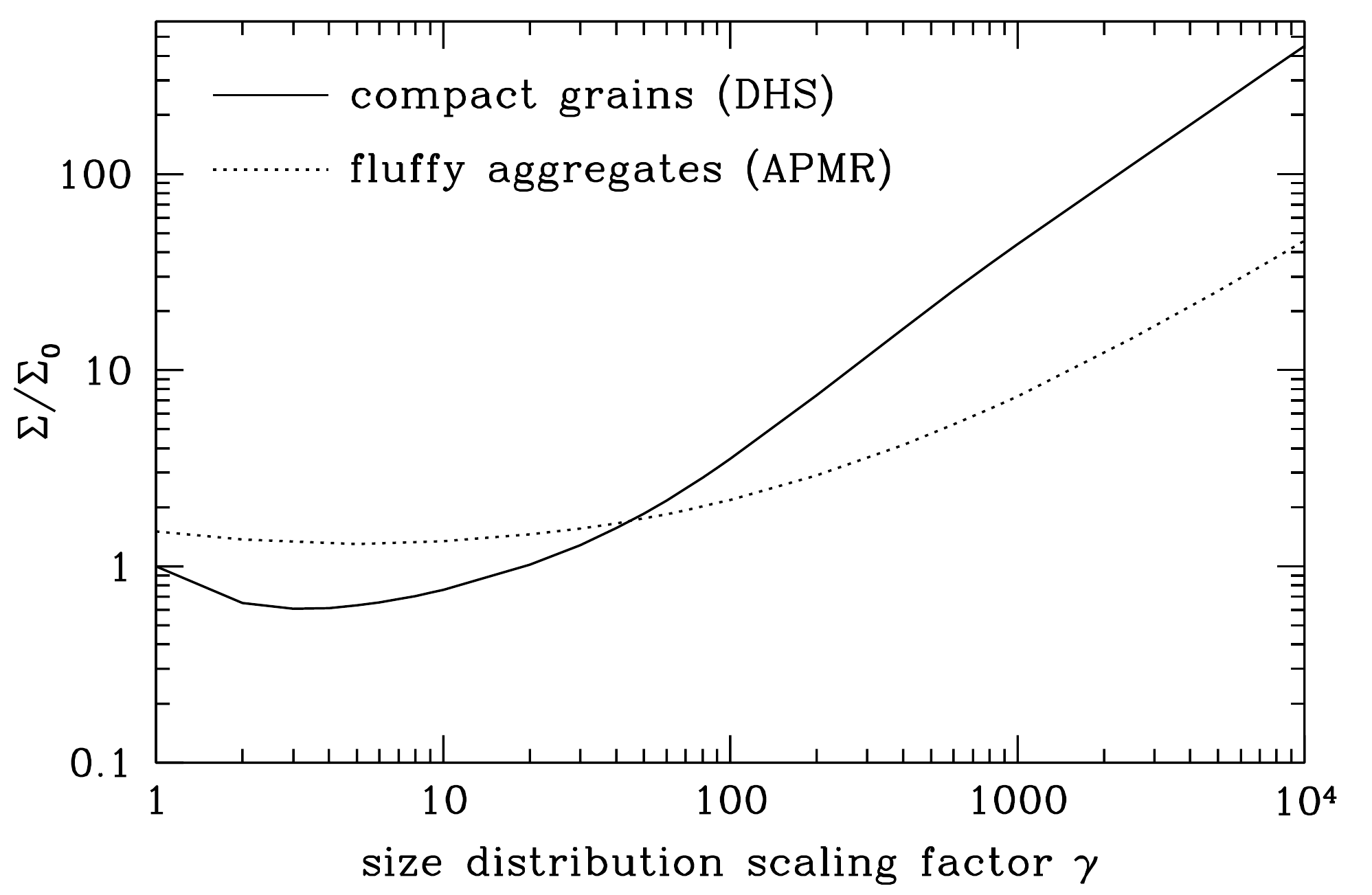}}}
\caption{The increase of the maximum allowed solid surface density as a function of the grain size distribution scaling factor $\gamma$ in the thermostat region of the disk. The solid curve is for the case of solid particle growth (computed using the Distribution of Hollow Spheres, DHS), the dotted curve is for the growth of fluffy aggregates (computed using the Aggregate Polarizability Mixing Rule, APMR).}
\label{fig:grain growth}
\end{figure}

The solid curve in Fig.~\ref{fig:grain growth} shows the increase in the maximum allowed solid surface density as a function of the scale factor $\gamma$. For example, we see that by increasing the grain size distribution by a factor 1000 the maximum allowed surface density increases by a factor $\sim$44. This is because at these grain sizes the Rosseland mean opacity (corrected for anisotropic scattering) is reduced by a factor of $\sim$44. The dotted curve in Fig.~\ref{fig:grain growth} shows the same but now for very fluffy aggregates. The opacities here are computed using the Aggregate Polarizability Mixing Rule \citep[APMR; see][]{2008A&A...489..135M}. It is clear that compared to the grain growth computed using compact grains (the solid curve) the fluffy aggregates have to grow to much larger sizes in order to reduce the opacity sufficiently to allow more solid mass to condense. If indeed sintering causes rigid bonds, prohibiting compaction of the formed aggregates, such fluffy aggregates are not an unlikely outcome of the aggregation process.

An interesting feature for the compact grains (and to a lesser extend for the fluffy aggregates) in Fig.~\ref{fig:grain growth} is that the opacity of the grains first increases, which causes more mass to evaporate. Only when the particles are sufficiently large the opacity decreases and the solid surface density can increase. This is caused by the increasing scattering efficiency when the grains grow from interstellar grain sizes to several microns in size. It implies that in order to form large particles, a barrier has to be taken. The complex behavior of grain growth under these conditions asks for a more detailed study. For the fluffy aggregates this effect is much weaker. This is because for these aggregated structures the scattering is heavily forward peaked which makes the effect of scattering on the radiative transfer very small.

Another possibility to decrease the opacity of the grains in the hot regions is to take away the species that contribute most to the opacity. We took a mixture of silicates, iron sulfide and carboneceous grains. It is well known that iron sulfide is only stable at temperatures below 680\,K \citep{1994ApJ...421..615P}. This means that the iron sulfide opacity source is most likely not there, though it can be argued that metallic iron will take over. Removing iron sulfide from the mixture we decrease the Rosseland mean opacity at 1500\,K by a factor $\sim$1.2, so this would allow only somewhat more mass to form. The dominant species of opacity in our computations is carbon. If we were to remove all the carboneceous species but leave the iron sulfide we reduce the Rosseland mean opacity at 1500\,K by a factor $\sim$3. Indeed, we find that the surface density in the inner region for our model without carboneceous solids is a factor of 3 higher that our standard model. Removing both the carboneceous species and the iron sulfide reduces the Rosseland mean opacity at 1500\,K by a factor $\sim$20. Though we did not perform the exact radiative transfer computations we can deduce that this would allow $\sim20$ times more mass in silicate grains to condense in the hot inner regions of the disk. This is assuming no other species adding opacity, like for instance metallic iron, form in these parts of the disk. Since metallic iron is a very strong opacity source (even slightly stronger than iron sulfide) even a small amount would already increase the opacity again to approximately the same level as with iron sulfide and carbon present.

In conclusion we find that the solid surface density in the region controlled by evaporation due to viscous heating is completely determined by the opacity of the grains and thus heavily depends on the local grain size and chemical composition.

\subsection{Coagulation timescale in the thermostat region}

Another important effect of the opacity-temperature control system is that the reduced number density of small dust grains will have a significant influence on the coagulation time scale in that region.  The time scale at which coagulation can proceed depends strongly on the initial number density of grains at which the aggregation process has to start.  Smaller particle densities increase the mean free path for particles and, consequently, the collision time for a dust particle to find another one to stick to and grow.  This effect is extended through the further growth phases.  \citet{2008A&A...491..663D} have shown that this effect extends the time scales for removing small grains from the disk by coagulation approximately inversely proportional to the initial number density.  This means that that coagulation will be strongly slowed down in this region, and that the time scales over which opacity dominates the number of grains that can be present in the thermostat region might be determined by the mass accretion time scale of the disk rather than the coagulation time scale.  We believe that this effect can have significant influence on coagulation and disk evolution models.  Further studies on this issue are necessary and will be the subject of a follow-up paper.

\section{Conclusions}

We successfully computed the solid density structure and snow line in the early Solar nebula for various parameters of the mass accretion rate and grain composition. This was done using full 3-D radiative transfer taking into account both the energy released by viscous heating as well as radiation from the Sun. The density structure of both the gas and the solids are computed self consistently using an $\alpha$ description for the viscosity and detailed evaporation physics for the dust and ice. We find that:
\begin{itemize}
\item The location of the snow line is determined by the opacity of the grains and the mass accretion rate. We show that approximate radiative transfer methods to compute the location of the snow line are quite accurate.
\item At high mass accretion rates the midplane temperature in the inner few AU can exceed the silicate evaporation temperature. This causes a balance between condensation and evaporation of matter acting like a thermostat keeping the midplane at the silicate evaporation temperature over a significant region. The solid mass in the inner regions is dramatically reduced by this effect and models computing the solid state surface density of the early Solar system should take this into account.
\item For our standard set of parameters the surface density in the solid state in the region of the terrestrial planets is below the MMSN for all values of the mass accretion rate. This poses serious constraints on the scenarios of planet formation. It could point to a significantly lower opacity value at the time of planet formation, which can be achieved by increasing the average grain size or significantly change the chemical composition.
\item At high accretion rates the increased midplane temperature of the disk due to viscous heating causes the midplane temperatures to be $>1000\,$K in a relatively large part of the disk. Under these conditions thermal processing of dust material as well as sintering of aggregates becomes important.
\item We find that for the high mass accretion rates the effects of convective cooling of the midplane are important to take into account.
\end{itemize}

\newpage

\newpage

\newpage

\newpage

\newpage

\newpage

\newpage


\begin{thebibliography}{42}
\expandafter\ifx\csname natexlab\endcsname\relax\def\natexlab#1{#1}\fi
\expandafter\ifx\csname url\endcsname\relax
  \def\url#1{\texttt{#1}}\fi
\expandafter\ifx\csname urlprefix\endcsname\relax\def\urlprefix{URL }\fi

\bibitem[{{Begemann} et~al.(1994){Begemann}, {Dorschner}, {Henning},
  {Mutschke}, and {Thamm}}]{1994ApJ...423L..71B}
{Begemann}, B., {Dorschner}, J., {Henning}, T., {Mutschke}, H., {Thamm}, E.,
  Mar. 1994. {A laboratory approach to the interstellar sulfide dust problem}.
  \apjl 423, L71--L74.

\bibitem[{{Bjorkman} and {Wood}(2001)}]{2001ApJ...554..615B}
{Bjorkman}, J.~E., {Wood}, K., Jun. 2001. {Radiative Equilibrium and
  Temperature Correction in Monte Carlo Radiation Transfer}. \apj 554,
  615--623.

\bibitem[{{Brauer} et~al.(2008){Brauer}, {Dullemond}, and
  {Henning}}]{2008A&A...480..859B}
{Brauer}, F., {Dullemond}, C.~P., {Henning}, T., Mar. 2008. {Coagulation,
  fragmentation and radial motion of solid particles in protoplanetary disks}.
  \aap 480, 859--877.

\bibitem[{{Chiang} and {Goldreich}(1997)}]{1997ApJ...490..368C}
{Chiang}, E.~I., {Goldreich}, P., Nov. 1997. {Spectral Energy Distributions of
  T Tauri Stars with Passive Circumstellar Disks}. \apj 490, 368--376.

\bibitem[{{Crida}(2009)}]{2009ApJ...698..606C}
{Crida}, A., Jun. 2009. {Minimum Mass Solar Nebulae and Planetary Migration}.
  \apj 698, 606--614.

\bibitem[{{D'Alessio} et~al.(1998){D'Alessio}, {Canto}, {Calvet}, and
  {Lizano}}]{1998ApJ...500..411D}
{D'Alessio}, P., {Canto}, J., {Calvet}, N., {Lizano}, S., Jun. 1998. {Accretion
  Disks around Young Objects. I. The Detailed Vertical Structure}. \apj 500,
  411--427.

\bibitem[{{Davis}(2005{\natexlab{a}})}]{2005ApJ...620..994D}
{Davis}, S.~S., Feb. 2005{\natexlab{a}}. {Condensation Front Migration in a
  Protoplanetary Nebula}. \apj 620, 994--1001.

\bibitem[{{Davis}(2005{\natexlab{b}})}]{2005ApJ...627L.153D}
{Davis}, S.~S., Jul. 2005{\natexlab{b}}. {The Surface Density Distribution in
  the Solar Nebula}. \apjl 627, L153--L155.

\bibitem[{{Desch}(2007)}]{2007ApJ...671..878D}
{Desch}, S.~J., Dec. 2007. {Mass Distribution and Planet Formation in the Solar
  Nebula}. \apj 671, 878--893.

\bibitem[{{Dodson-Robinson} et~al.(2009){Dodson-Robinson}, {Willacy},
  {Bodenheimer}, {Turner}, and {Beichman}}]{2009Icar..200..672D}
{Dodson-Robinson}, S.~E., {Willacy}, K., {Bodenheimer}, P., {Turner}, N.~J.,
  {Beichman}, C.~A., Apr. 2009. {Ice lines, planetesimal composition and solid
  surface density in the solar nebula}. Icarus 200, 672--693.

\bibitem[{{Dominik} and {Dullemond}(2008)}]{2008A&A...491..663D}
{Dominik}, C., {Dullemond}, C.~P., Dec. 2008. {Coagulation of small grains in
  disks: the influence of residual infall and initial small-grain content}.
  \aap 491, 663--670.

\bibitem[{{Dorschner} et~al.(1995){Dorschner}, {Begemann}, {Henning}, {J{\"
  a}ger}, and {Mutschke}}]{1995A&A...300..503D}
{Dorschner}, J., {Begemann}, B., {Henning}, T., {J{\" a}ger}, C., {Mutschke},
  H., Aug. 1995. {Steps toward interstellar silicate mineralogy. II. Study of
  Mg-Fe-silicate glasses of variable composition.} \aap 300, 503--520.

\bibitem[{{Dullemond} et~al.(2007){Dullemond}, {Hollenbach}, {Kamp}, and
  {D'Alessio}}]{2007prpl.conf..555D}
{Dullemond}, C.~P., {Hollenbach}, D., {Kamp}, I., {D'Alessio}, P., 2007.
  {Models of the Structure and Evolution of Protoplanetary Disks}. Protostars
  and Planets V, 555--572.

\bibitem[{{Fabian} et~al.(2000){Fabian}, {J{\" a}ger}, {Henning}, {Dorschner},
  and {Mutschke}}]{2000A&A...364..282F}
{Fabian}, D., {J{\" a}ger}, C., {Henning}, T., {Dorschner}, J., {Mutschke}, H.,
  Dec. 2000. {Steps toward interstellar silicate mineralogy. V. Thermal
  Evolution of Amorphous Magnesium Silicates and Silica}. \aap 364, 282--292.

\bibitem[{{Garaud} and {Lin}(2007)}]{2007ApJ...654..606G}
{Garaud}, P., {Lin}, D.~N.~C., Jan. 2007. {The Effect of Internal Dissipation
  and Surface Irradiation on the Structure of Disks and the Location of the
  Snow Line around Sun-like Stars}. \apj 654, 606--624.

\bibitem[{{Geiss}(1987)}]{1987A&A...187..859G}
{Geiss}, J., Nov. 1987. {Composition measurements and the history of cometary
  matter}. \aap 187, 859--866.

\bibitem[{{Gorti} and {Hollenbach}(2009)}]{2009ApJ...690.1539G}
{Gorti}, U., {Hollenbach}, D., Jan. 2009. {Photoevaporation of Circumstellar
  Disks By Far-Ultraviolet, Extreme-Ultraviolet and X-Ray Radiation from the
  Central Star}. \apj 690, 1539--1552.

\bibitem[{{Grevesse} and {Sauval}(1998)}]{1998SSRv...85..161G}
{Grevesse}, N., {Sauval}, A.~J., May 1998. {Standard Solar Composition}. Space
  Science Reviews 85, 161--174.

\bibitem[{{Hartmann} et~al.(2006){Hartmann}, {D'Alessio}, {Calvet}, and
  {Muzerolle}}]{2006ApJ...648..484H}
{Hartmann}, L., {D'Alessio}, P., {Calvet}, N., {Muzerolle}, J., Sep. 2006. {Why
  Do T Tauri Disks Accrete?} \apj 648, 484--490.

\bibitem[{{Hayashi}(1981)}]{1981IAUS...93..113H}
{Hayashi}, C., 1981. {Formation of the planets}. In: {D.~Sugimoto, D.~Q.~Lamb,
  \& D.~N.~Schramm} (Ed.), Fundamental Problems in the Theory of Stellar
  Evolution. Vol.~93 of IAU Symposium. pp. 113--126.

\bibitem[{{Henning} and {Stognienko}(1996)}]{1996A&A...311..291H}
{Henning}, T., {Stognienko}, R., Jul. 1996. {Dust opacities for protoplanetary
  accretion disks: influence of dust aggregates.} \aap 311, 291--303.

\bibitem[{{Hersant} et~al.(2001){Hersant}, {Gautier}, and
  {Hur{\'e}}}]{2001ApJ...554..391H}
{Hersant}, F., {Gautier}, D., {Hur{\'e}}, J., Jun. 2001. {A Two-dimensional
  Model for the Primordial Nebula Constrained by D/H Measurements in the Solar
  System: Implications for the Formation of Giant Planets}. \apj 554, 391--407.

\bibitem[{{Hubeny}(1990)}]{1990ApJ...351..632H}
{Hubeny}, I., Mar. 1990. {Vertical structure of accretion disks - A simplified
  analytical model}. \apj 351, 632--641.

\bibitem[{{Hur{\'e}}(2000)}]{2000A&A...358..378H}
{Hur{\'e}}, J., Jun. 2000. {On the transition to self-gravity in low mass AGN
  and YSO accretion discs}. \aap 358, 378--394.

\bibitem[{{Ida} and {Lin}(2008)}]{2008ApJ...685..584I}
{Ida}, S., {Lin}, D.~N.~C., Sep. 2008. {Toward a Deterministic Model of
  Planetary Formation. V. Accumulation Near the Ice Line and Super-Earths}.
  \apj 685, 584--595.

\bibitem[{{Ishimaru}(1978)}]{1978usaf.book.....I}
{Ishimaru}, A., 1978. {Wave propagation and scattering in random media. Volume
  I - Single scattering and transport theory}. Research supported by the
  U.S.~Air Force, NSF, and NIH.~New York, Academic Press, Inc.

\bibitem[{{Kama} et~al.(2009){Kama}, {Min}, and
  {Dominik}}]{2009A&A...506.1199K}
{Kama}, M., {Min}, M., {Dominik}, C., Nov. 2009. {The inner rim structures of
  protoplanetary discs}. \aap 506, 1199--1213.

\bibitem[{{Lecar} et~al.(2006){Lecar}, {Podolak}, {Sasselov}, and
  {Chiang}}]{2006ApJ...640.1115L}
{Lecar}, M., {Podolak}, M., {Sasselov}, D., {Chiang}, E., Apr. 2006. {On the
  Location of the Snow Line in a Protoplanetary Disk}. \apj 640, 1115--1118.

\bibitem[{{Mathis} et~al.(1977){Mathis}, {Rumpl}, and
  {Nordsieck}}]{1977ApJ...217..425M}
{Mathis}, J.~S., {Rumpl}, W., {Nordsieck}, K.~H., Oct. 1977. {The size
  distribution of interstellar grains}. \apj 217, 425--433.

\bibitem[{{Min} et~al.(2009){Min}, {Dullemond}, {Dominik}, {de Koter}, and
  {Hovenier}}]{2009A&A...497..155M}
{Min}, M., {Dullemond}, C.~P., {Dominik}, C., {de Koter}, A., {Hovenier},
  J.~W., Apr. 2009. {Radiative transfer in very optically thick circumstellar
  disks}. \aap 497, 155--166.

\bibitem[{{Min} et~al.(2005){Min}, {Hovenier}, and {de
  Koter}}]{2005A&A...432..909M}
{Min}, M., {Hovenier}, J.~W., {de Koter}, A., Mar. 2005. {Modeling optical
  properties of cosmic dust grains using a distribution of hollow spheres}.
  \aap 432, 909--920.

\bibitem[{{Min} et~al.(2008){Min}, {Hovenier}, {Waters}, and {de
  Koter}}]{2008A&A...489..135M}
{Min}, M., {Hovenier}, J.~W., {Waters}, L.~B.~F.~M., {de Koter}, A., Oct. 2008.
  {The infrared emission spectra of compositionally inhomogeneous aggregates
  composed of irregularly shaped constituents}. \aap 489, 135--141.

\bibitem[{{Mordasini} et~al.(2009){Mordasini}, {Alibert}, and
  {Benz}}]{2009A&A...501.1139M}
{Mordasini}, C., {Alibert}, Y., {Benz}, W., Jul. 2009. {Extrasolar planet
  population synthesis. I. Method, formation tracks, and mass-distance
  distribution}. \aap 501, 1139--1160.

\bibitem[{{Pollack} et~al.(1994){Pollack}, {Hollenbach}, {Beckwith},
  {Simonelli}, {Roush}, and {Fong}}]{1994ApJ...421..615P}
{Pollack}, J.~B., {Hollenbach}, D., {Beckwith}, S., {Simonelli}, D.~P.,
  {Roush}, T., {Fong}, W., Feb. 1994. {Composition and radiative properties of
  grains in molecular clouds and accretion disks}. \apj 421, 615--639.

\bibitem[{{Pollack} et~al.(1996){Pollack}, {Hubickyj}, {Bodenheimer},
  {Lissauer}, {Podolak}, and {Greenzweig}}]{1996Icar..124...62P}
{Pollack}, J.~B., {Hubickyj}, O., {Bodenheimer}, P., {Lissauer}, J.~J.,
  {Podolak}, M., {Greenzweig}, Y., Nov. 1996. {Formation of the Giant Planets
  by Concurrent Accretion of Solids and Gas}. Icarus 124, 62--85.

\bibitem[{{Poppe}(2003)}]{2003Icar..164..139P}
{Poppe}, T., Jul. 2003. {Sintering of highly porous silica-particle samples:
  analogues of early Solar-System aggregates}. Icarus 164, 139--148.

\bibitem[{{Preibisch} et~al.(1993){Preibisch}, {Ossenkopf}, {Yorke}, and
  {Henning}}]{1993A&A...279..577P}
{Preibisch}, T., {Ossenkopf}, V., {Yorke}, H.~W., {Henning}, T., Nov. 1993.
  {The influence of ice-coated grains on protostellar spectra}. \aap 279,
  577--588.

\bibitem[{{Ruden} and {Lin}(1986)}]{1986ApJ...308..883R}
{Ruden}, S.~P., {Lin}, D.~N.~C., Sep. 1986. {The global evolution of the
  primordial solar nebula}. \apj 308, 883--901.

\bibitem[{{Shakura} and {Sunyaev}(1973)}]{1973A&A....24..337S}
{Shakura}, N.~I., {Sunyaev}, R.~A., 1973. {Black holes in binary systems.
  Observational appearance.} \aap 24, 337--355.

\bibitem[{{Thommes} and {Duncan}(2006)}]{2006plfo.book..129T}
{Thommes}, E.~W., {Duncan}, M.~J., May 2006. {The accretion of giant-planet
  cores}. Cambridge University Press, pp. 129--146.

\bibitem[{{Warren}(1984)}]{1984ApOpt..23.1206W}
{Warren}, S.~G., Apr. 1984. {Optical constants of ice from the ultraviolet to
  the microwave}. \ao 23, 1206--1225.

\bibitem[{{Weidenschilling}(1977)}]{1977Ap&SS..51..153W}
{Weidenschilling}, S.~J., Sep. 1977. {The distribution of mass in the planetary
  system and solar nebula}. \apss 51, 153--158.

\end{thebibliography}
\end{document}